\documentclass[sigconf]{acmart} 

\AtBeginDocument{%
  \providecommand\BibTeX{{%
    \normalfont B\kern-0.5em{\scshape i\kern-0.25em b}\kern-0.8em\TeX}}}

\copyrightyear{2024}
\acmYear{2024}
\setcopyright{rightsretained}
\acmConference[CHI '24]{Proceedings of the CHI Conference on Human Factors in Computing Systems}{May 11--16, 2024}{Honolulu, HI, USA}
\acmBooktitle{Proceedings of the CHI Conference on Human Factors in Computing Systems (CHI '24), May 11--16, 2024, Honolulu, HI, USA}
\acmDOI{10.1145/3613904.3642752}
\acmISBN{979-8-4007-0330-0/24/05}

\usepackage{multirow}
\usepackage{hyperref}
\usepackage{subcaption}
\usepackage{listings}
\usepackage{algorithm}
\usepackage{algorithmic}
\usepackage{makecell}

\usepackage[english]{babel}

\newcommand{\pointone}[1]{\textcolor{black}{#1}}
\newcommand{\pointonecolor}[1]{\textcolor{black}{{#1}}}
\newcommand{\pointtwo}[1]{\textcolor{black}{#1}}
\newcommand{\pointtwocolor}[1]{\textcolor{black}{#1}}
\newcommand{\pointthree}[1]{\textcolor{black}{#1}}

\newcommand{\pointfour}[1]{\textcolor{black}{#1}}
\newcommand{\pointfive}[1]{\textcolor{black}{#1}}
\newcommand{\pointsix}[1]{\textcolor{black}{#1}}
\newcommand{\pointsixcolor}[1]{\textcolor{black}{#1}}
\newcommand{\pointseven}[1]{\textcolor{black}{#1}}
\newcommand{\pointsevencolor}[1]{\textcolor{black}{#1}}

\newcommand{\mr}[1]{\textcolor{black}{{#1}}} % minor? teal
\newcommand{\rr}[1]{\textcolor{black}{{#1}}}
\newcommand{\ar}[1]{\textcolor{black}{{#1}}} % violet
\newcommand{\sr}[1]{\textcolor{black}{{#1}}} % violet
\newcommand{\nr}[1]{\textcolor{black}{{#1}}} % violet

\begin{document}

% Chatorials / 
\title{\emph{AQuA}: Automated Question-Answering in Software Tutorial Videos with Visual Anchors 
}
% Using Generative Visual-Language Models 

%
% The "author" command and its associated commands are used to define
% the authors and their affiliations.
% Of note is the shared affiliation of the first two authors, and the
% "authornote" and "authornotemark" commands
% used to denote shared contribution to the research.
\author{Saelyne Yang}
\email{saelyne@kaist.ac.kr}
\additionalaffiliation{%
  \institution{KAIST}
  \city{Daejeon}
  \country{Republic of Korea}
}
\affiliation{%
  \institution{Autodesk Research}
  \city{Toronto}
  \state{Ontario}
  \country{Canada}
}

\author{Jo Vermeulen}
\email{jo.vermeulen@autodesk.com}
\affiliation{%
  \institution{Autodesk Research}
  \city{Toronto}
  \state{Ontario}
  \country{Canada}
}

\author{George Fitzmaurice}
\email{george.fitzmaurice@autodesk.com}
\affiliation{%
  \institution{Autodesk Research}
  \city{Toronto}
  \state{Ontario}
  \country{Canada}
}

\author{Justin Matejka}
\email{justin.matejka@autodesk.com}
\affiliation{%
  \institution{Autodesk Research}
  \city{Toronto}
  \state{Ontario}
  \country{Canada}
}



\begin{abstract}
Tutorial videos are a popular help source for learning feature-rich software. However, getting quick answers to questions about tutorial videos is difficult. We present an automated approach for responding to tutorial questions. By analyzing 633 questions found in 5,944 video comments, we identified different question types and observed that users frequently described parts of the video in questions. We then asked participants (N=24) to watch tutorial videos and ask questions while annotating the video with relevant visual anchors. Most visual anchors referred to UI elements and the application workspace. Based on these insights, we built \emph{AQuA}, a pipeline that generates useful answers to questions with visual anchors. We demonstrate this for Fusion 360, showing that we can recognize UI elements in visual anchors and generate answers using GPT-4 augmented with that visual information and software documentation. An evaluation study (N=16) demonstrates that our approach provides better answers than baseline methods.
\end{abstract}






\begin{teaserfigure}
    \centering
    \includegraphics[width=\linewidth]{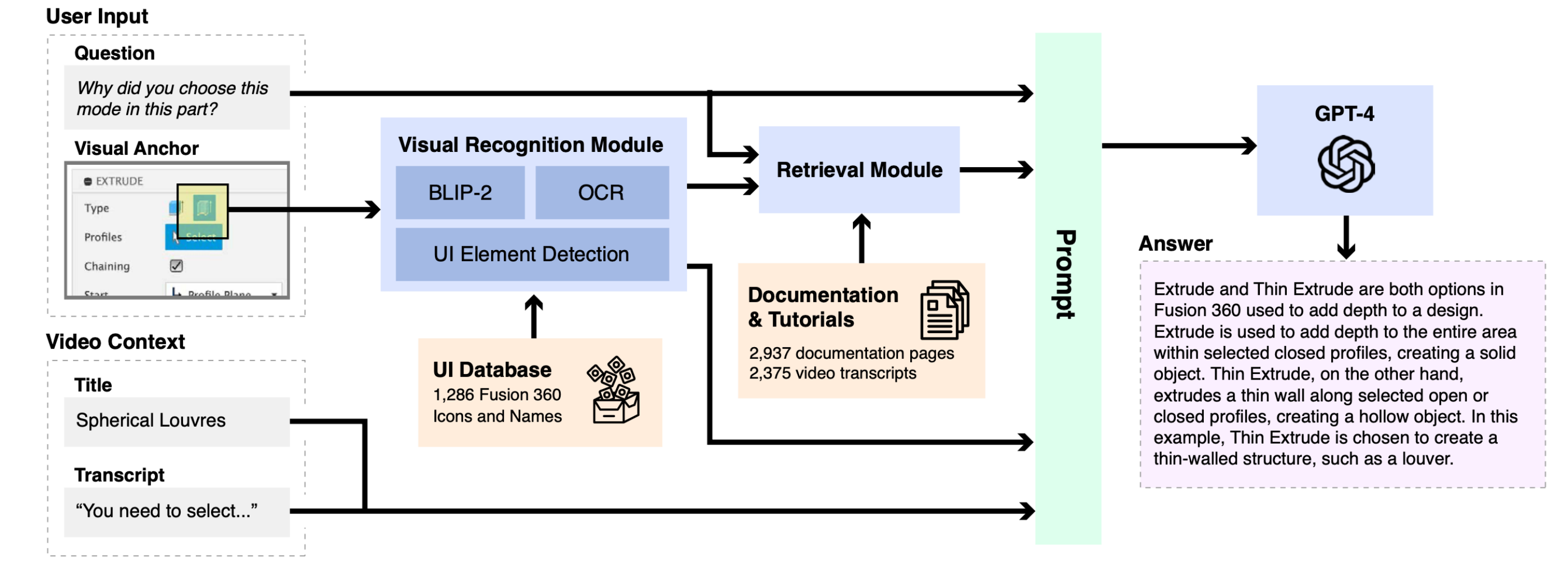}
    \caption{Overall architecture of our question-answer pipeline \emph{AQuA}, which generates useful responses to questions made in software tutorial videos. Questions are accompanied by visual anchors, which are specific visual elements of interest in the video. The Visual Recognition Module generates a textual description of the visual anchor. Combining the description with the question, the Retrieval Module retrieves relevant articles to the queries. Resources in yellow boxes are software-specific materials (in this case, for Fusion 360). Along with these retrieved articles, the question text, and the visual anchor description, we include the title and relevant transcript sentences of the tutorial video and feed them into GPT-4 through crafted prompts.}
    \label{fig:pipeline}
    \Description{A diagram that shows the overall architecture of our question-answer pipeline, AQuA. User Input includes Question and Visual Anchor, and Video Context includes Title and Transcript. Visual Recognition Module includes BLIP-2, OCR, and UI Element Detection.}
% \vspace*{6pt}
\end{teaserfigure}

\maketitle
\section{Introduction}

Tutorial videos are a popular resource for people seeking to learn how to use feature-rich software applications~\cite{one2019, pause2011, ambient2011}. These videos present workflows in great detail, with authors sharing their screens and often supplementing the workflow with verbal explanations.
However, despite the abundant information they provide, viewers can face difficulties in understanding or following the content~\cite{yang2020hard, chang2019howto}.
\pointthree{To gain a better understanding of the material or clarify uncertain segments, viewers might rewind to a specific position and rewatch it or skip forward to anticipate the next steps~\cite{chang2019howto, zhao2022jiggling, yang2022softvideo}. To streamline this process, a number of systems have been proposed to help users navigate videos based on their current context and inquiry ~\cite{chang2021rubyslippers, yang2022softvideo}.}

\pointthree{However, there are instances where users still struggle to comprehend certain parts even after jumping around in the video, especially if the video doesn't address their specific queries. In such cases,}
they often leave questions in the comments section, requesting further explanations about specific parts of the video~\cite{codingcomment}. 
While timely answers to questions are crucial for effective learning from tutorials, obtaining answers from the community or the tutorial authors can take hours or days. In some instances, questions may even go unanswered. This delay in addressing questions can disrupt the learning process and discourage viewers from fully engaging with tutorial content.

To address the problem, we explored methods to automate the process of answering questions about tutorial videos. We first begin with an in-depth analysis of user question-asking behavior. To gain insights into this behavior, we collected a dataset of all 5,944 comments from the top 20 most popular video tutorials for Autodesk Fusion 360, a 3D Computer-Aided Design (CAD) software application. After identifying 663 questions in the comments, we further identified \sr{four main categories of questions}: questions about the tutorial content \sr{(`Content')}, questions regarding learners' personal settings or challenges \sr{in regard to the tutorial (`User')}, questions concerning the video's meta-information \sr{(`Meta')}, and questions not directly related to the content \sr{(`General')}.

\pointone{We decided to focus on the first two categories due to their relatedness to the tutorial content (i.e., `Content' and `User').} A notable pattern that emerged in these categories was the tendency of users to reference parts of the video in their comments to provide temporal and spatial context. 
\pointtwo{In particular, we noticed several cases of referring to the visual part of the video, which aligns with previous findings on general videos~\cite{yarmand2019reference}. In contrast to this prior work~\cite{yarmand2019reference}, we observed this practice in software tutorial videos, in which it can be} particularly evident since they feature visual demonstrations through screen sharing.

Inspired by these findings, our research delves further into the types of visual content that users reference in their questions. To explore this, we developed a system that allows users to ask questions by creating visual anchors, which are specific visual elements of interest in the video that pertain to their questions. We conducted a study with 24 participants where they were asked to watch a tutorial video and formulate questions with relevant visual anchors. We selected four tutorial videos each for three software applications – Fusion 360, Photoshop, and Excel, resulting in a total of 12 tutorial videos. In this study, we collected 217 questions, each accompanied by one or more visual anchors relevant to the question.

Our analysis showed that the majority of visual anchors were related to specific user interface (UI) elements and the workspace of the software applications. Furthermore, nearly half of the questions required these visual anchors to supply essential contextual information. These findings underline the critical role of visual context in comprehending and responding to user queries in tutorial videos.

Based on our findings, we developed \emph{AQuA}, a comprehensive pipeline to generate useful responses to questions that include visual anchors. Developed specifically for Fusion 360 as a case study, our pipeline identifies software UI elements in the visual anchors associated with questions and generates responses by leveraging the Large Language Model (LLM) GPT-4~\cite{openai2023gpt4}, which is further enriched with specific knowledge about the software. We achieve this by drawing on official documentation and tutorial resources, which are generally available for most software applications.

We then evaluated our pipeline in a study with 16 Fusion 360 users. The results demonstrate that our pipeline produces more correct and helpful answers compared to baseline methods, and was the most favored. In the discussion, we outline design considerations for question-answering systems, providing insights into interactive and responsive learning experiences within the context of tutorial videos.

In summary, this paper presents the following main contributions:
\begin{itemize}
    \item {Two formative studies that uncover users' question-asking behavior in software tutorial videos.}
    \item{A comprehensive pipeline \emph{AQuA}, \pointfive{which takes a novel multimodal approach with visual recognition and LLMs augmented with software-specific materials} to generate answers to tutorial questions with visual anchors.}
    \item{An evaluation study that demonstrates the effectiveness of our pipeline in addressing user questions.}
\end{itemize}

% an augmented LLM and visual recognition in a novel multimodal approach
% an LLM-based multimodal pipeline augmented with software-specific materials and visual recognition information
\section{Related Work}
Our work presents an automated approach for answering questions asked on software tutorial videos that reference specific elements within the video. We discuss related work in the areas of software learning \pointseven{and UI understanding, video navigation and control, video question answering, and referencing techniques.}

\subsection{\pointsevencolor{Software Learning, Tutorials, and UI Understanding}}
Software applications such as Adobe Photoshop and Autodesk AutoCAD provide rich functionality to accommodate users working on a wide range of tasks. However, it can be challenging to learn how to use such \textit{feature-rich software}. Previous research has explored ways to simplify this learning process. For instance, Masson et al.~\cite{trialerror} have focused on a ``learning-by-doing'' approach, introducing techniques that make users' trial-and-error cycle more meaningful. However, while users can learn some aspects of the software by themselves, they may also encounter challenges and need to seek out additional help. A line of work has been conducted to support help-seeking behavior by facilitating searching for functionality, such as by using screenshots of software with Sikuli~\cite{sikuli} or through multimodal input that refers to specific elements in the software in ReMap~\cite{remap}. Other research has presented methods for offering contextual help, such as presenting web pages or videos in AmbientHelp~\cite{ambient2011}, or specific segments within videos with RePlay~\cite{replay2019}. Furthermore, researchers explored ways to better connect software users to their peers in the community. IP-QAT~\cite{ipqat} enabled users to post questions directly within the software, while MicroMentor~\cite{micromentor} facilitated getting help from experienced software users in real time. 

Another way users learn about software is through tutorials. A body of work has focused on improving the usability of tutorials. Efforts have been made to enhance navigation to relevant parts of tutorial videos. Waken~\cite{waken2012} recognizes and displays information about tools used in the tutorials, while other research such as ToolScape~\cite{toolscape2014} and Fraser et al.~\cite{livestreamSegment2020} segment the videos into sections for easier navigation. Some systems have integrated the user's workflow into the video. Examples include SoftVideo~\cite{yang2022softvideo}, which provides real-time feedback on progress with tutorial content, Pause-and-Play~\cite{pause2011} which controls video playback based on user progress, and Nguyen and Liu's work that allows users to learn directly from the video as if they are interacting with the software itself~\cite{responsive2015}. 

\pointseven{
Beyond desktop applications, research has explored mobile and web applications, focusing on UI understanding for tasks like screen summarization and task automation. Using datasets like RICO~\cite{rico} and WebUI~\cite{webui}, a number of approaches have leveraged view hierarchy information of screens. For instance, Screen2Vec~\cite{screen2vec} transforms UI screens into embeddings for tasks like screen retrieval, and Screen2Words~\cite{screen2words} generates a summary of information that a UI screen contains. Combined with Large Language Models, Wang et al.~\cite{conversationMobileUI} proposed an approach that enables conversational interaction with mobile UI. Recently, Spotlight ~\cite{Li2022SpotlightMU} has been proposed, which does not require a view hierarchy but relies solely on visual screenshots to generate textual descriptions. In the realm of pixel understanding, Chen et al. ~\cite{chen22icon} focus on detecting icon types, while Zhang et al. ~\cite{zhang21screen} focus on detecting UI element types, which has contributed to improved accessibility in mobile applications. 
% To better understand users' questions on software videos, 
In our work,
we go beyond identifying the type of software UI elements in a single static image. We also identify the specific name of software UI elements in visual anchors that are captured and cropped in videos, by constructing a UI image database for a particular software application. Building on UI element understanding, our work aims to offer direct help while users learn from software tutorial videos by addressing their questions.}

\subsection{\pointsevencolor{Video Navigation and Control}}
\pointthree{\pointseven{
Users often encounter challenges when engaging with tutorial videos, struggling to comprehend or follow the content~\cite{yang2020hard, chang2019howto}. In these situations, users may seek specific segments within the video to address challenges or resolve confusion~\cite{chang2019howto, zhao2022jiggling, yang2022softvideo}. To facilitate the process of locating needed segments in videos, several researchers have proposed approaches to organize video content in a structured way~\cite{toolscape2014, livestreamSegment2020, anh21makeup, yang23beyond}. For example, Yang et al.~\cite{yang23beyond} have demonstrated that displaying information types for each segment in how-to videos can enhance the search for answers within the video content. In efforts to enhance users' direct control over videos, various studies have explored how users can interact with videos conversationally~\cite{chang2019howto, zhao2022jiggling, lin2023multimodal}. For instance, Rubyslippers~\cite{chang2021rubyslippers} allows content-based navigation of how-to videos using voice commands. These approaches empower users to pinpoint specific points of interest or points they need in their current context. However, there are still instances where users' queries or needs go beyond the information presented in the tutorial~\cite{Lafreniere_Bunt_Lount_Terry_2021}.
In our work, we address these cases where what users seek is not present in the video itself, by leveraging software-specific materials and the wealth of knowledge in Large Language Models. 
}}

\subsection{Video Question Answering}
Asking questions about the video content is a common user behavior~\cite{Madden2013ACS, codingcomment, pokharel2021classifying}. Users often ask questions about parts of the video that need further explanation or request additional content~\cite{codingcomment}.
% \rr{In such cases, users pose questions about the video content~\cite{Madden2013ACS, codingcomment, pokharel2021classifying}, often seeking further explanation or requesting additional content~\cite{codingcomment}.} 
Previous research in HCI has developed systems for question-answering in specialized domains such as programming~\cite{programmingQA}, math \cite{mathbot}, and children's general knowledge questions~\cite{dapie}. GVQA~\cite{song23graph} and Kim et al.~\cite{kim2023chart} have explored chart and graph comprehension through question-answering.

To address video question answering, the Computer Vision (CV) and Natural language processing (NLP) communities have introduced computational approaches and datasets. 
\mr{Among them, several benchmark datasets have been introduced that focus on how-to videos. For example, HowToVQA69M~\cite{Yang2020JustAL} contains question-answer pairs that are automatically generated from transcribed narrations. On the other hand, How2QA~\cite{Li2020HeroHE} and iVQA~\cite{Yang2020JustAL} collected questions and answers by presenting videos to crowd workers.}
% Notable datasets such as HowToVQA69M~\cite{Yang2020JustAL}, iVQA~\cite{Yang2020JustAL}, and How2QA~\cite{Li2020HeroHE} include questions and corresponding answers related to how-to videos. 
\mr{In particular, TutorialVQA~\cite{Colas2019TutorialVQAQA} and PsTuts-VQA~\cite{zhao2021photoshop} focus on \emph{software tutorial videos}, collecting questions from crowd workers by presenting answer segments or having software experts craft questions. 
% However, the questions in these datasets are either artificially generated by crowd workers who are presented with answer segments, or they are crafted by software experts. 
However, since these questions are artificially generated or automatically generated from transcripts,} using these questions can be limiting when developing approaches to address questions from real-world users. Additionally, unlike these datasets, our approach goes beyond text-only queries by incorporating associated visual elements, which reflects how users would naturally ask questions. 

% HowToVQA69M: automatic generation from transcripts / iVQA: crowd workers -> answerable from video / How2QA: crowd workers / TutorialVQA: presenting answer segments to crowdsourced workers and soliciting question / PsTuts-VQA: domain experts

% In summary, our work not only investigates the question-asking behavior of users in the context of tutorial videos but also thoughtfully designs an answer pipeline based on these insights. We believe this provides valuable insights into how video question-answering systems should be designed.

In summary, our work takes a step further in assisting users with software tutorial videos by offering quick, automated and accurate responses to questions. Unlike previous video question-answering systems, our approach handles questions accompanied by visual elements, reflecting a common pattern in users' question-asking behavior on tutorial videos.

\begin{table*}[h]
\centering
\begin{tabular}{c|c|p{50mm}|p{53mm}}
\toprule

{\textbf{Category}}&{\textbf{Type}} & \hfil \textbf{Definition} & \hfil \textbf{Example} \\ \hline

% \multirow{4}{*}{\vtop{\hbox{\strut Content }\hbox{\strut (N=251)}}} 
\multirow{4}{*}{Content} 
& \multicolumn{1}{c|}{Concept} & {Asks about specific concepts explained in the video} & {\textit{``Can you explain the difference between Press Pull and Extrude?''}} \\ \cline{2-4}
\multicolumn{1}{c|}{} & \multicolumn{1}{c|}{Method} & {Seeks clarification about a particular action or process} & {\textit{``How to import an existing path (e.g., from an SVG file or a Fusion sketch) into an electronics design?''}} \\
\cline{2-4}
\multicolumn{1}{c|}{} &
\multicolumn{1}{c|}{Reason} & {Asks the rationale behind certain instructions} & {\textit{``Any reason for not using a construction line for the middle line?''}} \\ \cline{2-4}
\multicolumn{1}{c|}{} &
\multicolumn{1}{c|}{Alternative} & {Explores other ways to accomplish a task} & {\textit{``Would the Intersect command work the same way for that?''}} \\
\cline{2-4}
\hline

% \multirow{2}{*}{\vtop{\hbox{\strut User }\hbox{\strut (N=144)}}}  
\multirow{2}{*}{User}  
& \multicolumn{1}{c|}{Problem} & {Reports issues encountered while following the tutorials} & {\textit{``Anyone else have issues with slow-downs when using several rectangular or circular arrays in sketches?''}} \\ \cline{2-4}
\multicolumn{1}{c|}{} & \multicolumn{1}{c|}{Advice} & {Seeks personalized tips or guidance} & {\textit{``I have a specific shape board where some components have to be located to fit in set openings. Advice?''}} \\ \hline 

% \multirow{3}{*}{\vtop{\hbox{\strut Meta }\hbox{\strut (N=106)}}} 
\multirow{3}{*}{Meta} 
& \multicolumn{1}{c|}{Content} & {Asks about the topic or duration of the tutorial} & {\textit{``What's today's topic? How long is this stream going on?''}} \\ \cline{2-4}
\multicolumn{1}{c|}{} & \multicolumn{1}{c|}{Setting} & {Asks about the technical details of the tutorial production} & {\textit{``What software are you using to screencast?''}} \\
\cline{2-4}
\multicolumn{1}{c|}{} &
\multicolumn{1}{c|}{Resource} & {Requests materials used in the tutorial} & {\textit{``I was wondering where I can get the reference images?''}} \\ 
\hline

% \multirow{3}{*}{\vtop{\hbox{\strut General }\hbox{\strut (N=132)}}}  
\multirow{3}{*}{General}  
& \multicolumn{1}{c|}{Software} & {Asks about the software’s features} & {\textit{``Is everything going to be migrated into Fusion in the future?''}} \\ \cline{2-4}
\multicolumn{1}{c|}{} & \multicolumn{1}{c|}{Future Content} & {Suggests topics to be covered in upcoming tutorials} & {\textit{``I hope in the future you will also make a real-time simulator in between circuit design and board design.''}} \\
\cline{2-4}
\multicolumn{1}{c|}{} &
\multicolumn{1}{c|}{Others} & {Asks about general topics} & {\textit{``Is there a WhatsApp community for Fusion 360?''}} \\ 
% \hline

\bottomrule
\end{tabular}
\caption{Definition and examples of question categories and types derived from Formative Study 1. Minor grammar errors and typos in example comments are corrected.}
\Description{Table that shows Category, Type, Definition, and Example of questions.}
\label{table:defintion}
\end{table*}

\subsection{Video Referencing}
% Another relevant thread in video interaction is...
Referencing specific audio or visual content within a video is a common practice during video interactions~\cite{yarmand2019reference, Schultes2013LeaveAC}. Yarmand et al. have explored referencing patterns in user comments on videos, identifying types of references such as object, speech, and concept references~\cite{yarmand2019reference}. While traditional video interfaces offer limited support for references (usually restricted to timestamps), this work suggests that the ability to easily refer to a part of a video enables a range of different applications\mr{, including enhanced engagement in live streams~\cite{snapstream, vispoll}.}

The ability to refer to parts of a video fosters a clear understanding of what others are discussing and facilitates pinpointed feedback or areas of confusion. Mudslide~\cite{mudslide} has shown that spatially contextualizing students' confusion points on lecture slides can benefit both learners and instructors. HyperButton~\cite{hyperbutton} has demonstrated that questions and answers anchored to specific frames can serve as valuable resources for future learners. As shown in Korero~\cite{korero}, this can also enhance mutual understanding among users and facilitate rich discussions about the video content. 
\mr{Video referencing can also enhance the learning experience. VideoSticker~\cite{videosticker} allows users to extract specific objects from videos, helping learners take notes of the video content. Nguyen and Liu have introduced a tutorial video system where users can directly interact with videos by clicking on software elements in screencast videos~\cite{responsive2015}.}
% These advantages are not confined to pre-recorded, asynchronous videos; they also enhance engagement in live streams~\cite{snapstream, vispoll}. 
In our research, we investigate video referencing behavior within the context of question-asking. This enables users to articulate their questions more effectively by making direct visual references to specific portions of the video.

\section{Formative Study 1: Software Tutorial Video Question Analysis}

\pointtwo{To get insights into the requirements for the answer pipeline, we conducted two series of formative studies to understand users' question-asking behavior in software tutorial videos. In our first study, we aimed to 
understand the types of questions users ask and identify the information required to provide answers.} To achieve this, we gathered user comments from YouTube, focusing on Autodesk Fusion 360---a widely used and feature-rich 3D CAD software application---as our case study.

\begin{figure*}[t]
    \centering
    \includegraphics[width=0.7\linewidth]{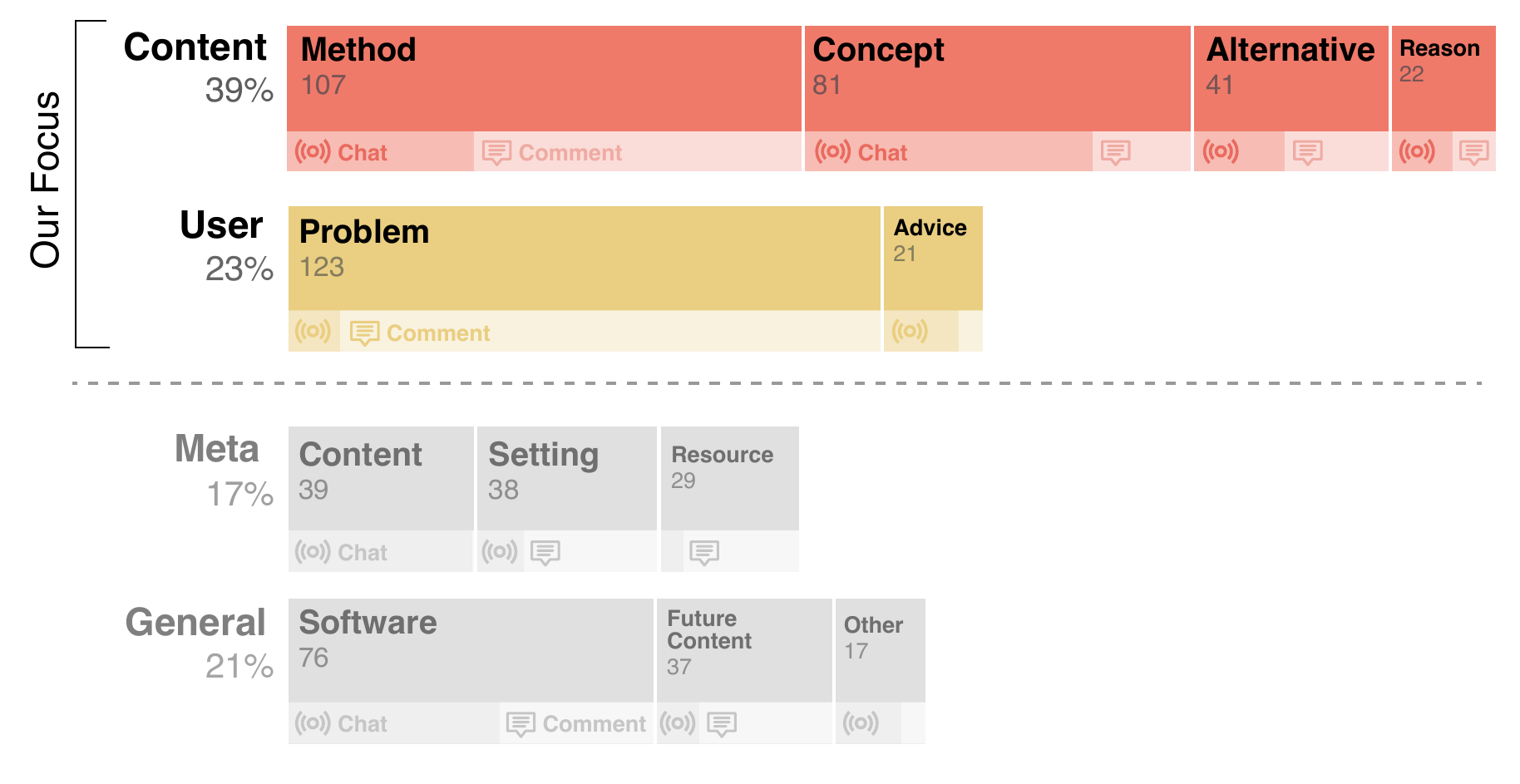}
    \caption{\pointone{Categories and Types of questions identified from the analysis. Each row represents a category and each block represents a type. Under each block, the areas on the left and right represent live chat and comment data, respectively. Our focus is on \textit{Content} and \textit{User} questions, as these are vital for comprehending the tutorial and can often be answered without the involvement of the tutorial authors or software vendor.}}
    \label{fig:type}
    \Description{Visualization of 12 types under 4 categories. Each row represents a category, and each block represents a type. Under each block, the areas on the left and right represent chat and comment data, respectively.}
% \vspace*{6pt}
\end{figure*}

\subsection{\pointtwocolor{Method}}
We selected the top 20 popular archived live streams\footnote{\mr{\url{www.youtube.com/@adskFusion/streams}}} from Fusion 360's official YouTube channel for our analysis. 
\pointfour{These live streams are instructional videos that are created by the official channel with the purpose of explaining how to perform tasks or sharing tips with Fusion 360 learners.}
We chose archived live streams as opposed to non-live tutorial videos as the live streams allowed us to investigate question-asking behavior in a comprehensive way, considering both comments made during real-time viewing where immediate help may be available, and comments made during asynchronous viewing after the live stream had ended, where \textit{live} help is not accessible. We gathered a total of 5,944 messages, which included 3,905 live chat messages sent during the streams and 2,039 comments posted on the same archived videos. We used Chat Downloader~\cite{chatdata-api} to collect live chat messages and used the YouTube Data API~\cite{ytdata-api} to gather comment data.
%\footnote{\pointfour{Our use of the API complies with the YouTube API Services Terms of Service.}}.

We examined the collected comments to get a sense of the types of questions that were asked. We first filtered for comments that were (1) asking questions or making requests, (2) posted by viewers (not by the tutorial author or moderator), and (3) initial comments (not replies). This resulted in 633 questions out of the 5944 comments. 
% The authors coded each question and iteratively refined the schema for categorizing the types of questions.
\pointfour{Similar to Yarmand et al.~\cite{yarmand2019reference}, the lead author performed thematic coding of the set of 633 questions and iteratively discussed with the other authors to validate the codes and resolve any conflicts. After finalizing the codes, we grouped them into four main categories, reflecting the overarching themes of the questions.}

% \begin{figure*}[t]
%     \centering
%     \includegraphics[width=0.8\linewidth]{figure/TutorialBot.pdf}
%     \caption{\pointone{Categories and Types of questions identified from the analysis. \emph{AQuA}'s focus is on Content and User questions, as these are vital for comprehending the tutorial and can often be answered without the involvement of the tutorial authors or software vendor. Please refer to Appendix~\ref{sec:taxonomy} for detailed explanations of each type.}}
%     \label{fig:focus}
%     \Description{Visualization of 12 types under 4 categories.}
% % \vspace*{6pt}
% \end{figure*}

\subsection{\pointtwocolor{Results}}

% Below, we describe the 12 distinct types of questions we identified, which are grouped under four main categories: Content, User, Meta, and General. Figure~\ref{fig:type} illustrates the distribution of these types for live chat and comment data.

Table~\ref{table:defintion} provides definitions and examples for the 12 distinct types of questions we identified, which are divided into under four main categories: 
% Fiure~\ref{fig:focus} describes the four main categories that emerged from our analysis:
\begin{itemize}
    \item \textbf{Content (39.7\%):} questions about the tutorial content presented in the tutorial.
    \item \textbf{User (22.6\%):} questions about the viewer’s settings or challenges \rr{in regard to the tutorial}.
    \item \textbf{Meta (16.7\%):} questions about the tutorial video’s meta-information.
    \item \textbf{General (20.9\%):} questions that are not directly related to the tutorial content.
\end{itemize}

% Figure~\ref{fig:type} illustrates the distribution of these types of live chat and comment data. 
\pointtwo{
The \textbf{`Content'} category encompasses questions related to the tutorial content presented in the tutorial, such as questions about concepts explained in the video (e.g., \textit{``Can you explain the difference between Press Pull and Extrude?''}) or questions that ask about the rationale behind certain instructions (e.g., \textit{``Is there a reason for not using a construction line for the middle line?''}).
\textbf{`User'} focuses on questions about the viewer's personal settings or challenges, such as questions reporting issues they encounter (e.g., \textit{``That marking menu doesn't seem to work. How do I fix it?''}) or those seeking tips or guidance (e.g., \textit{``I have a specific shape board where some components have to be located to fit in set openings. Advice?''}).
\textbf{`Meta'} focuses on questions about the tutorial video's meta-information, such as the technical details of the tutorial production (e.g., \textit{``What software are you using to screencast?''}) or materials used in the tutorial (e.g., \textit{``I was wondering where I can get the reference images?''}).
Finally, \textbf{General} includes questions that are not directly related to the tutorial content, such as those asking about the software's features (\textit{``Is everything going to be migrated into Fusion in the future?''}).}

\pointtwocolor{
Overall, \textbf{`Content'} and \textbf{`User'} questions are related to the tutorial content, seeking comprehension or practical help in understanding the tutorial. In contrast, \textbf{`Meta'} and \textbf{`General'} questions concern the meta-information or information unrelated to the tutorial's core content. These inquiries typically require insights from either the tutorial author (e.g., providing material resources) or software developers (e.g., detailing new feature timelines).}

\begin{figure*}[t]
    \centering
    \includegraphics[width=\linewidth]{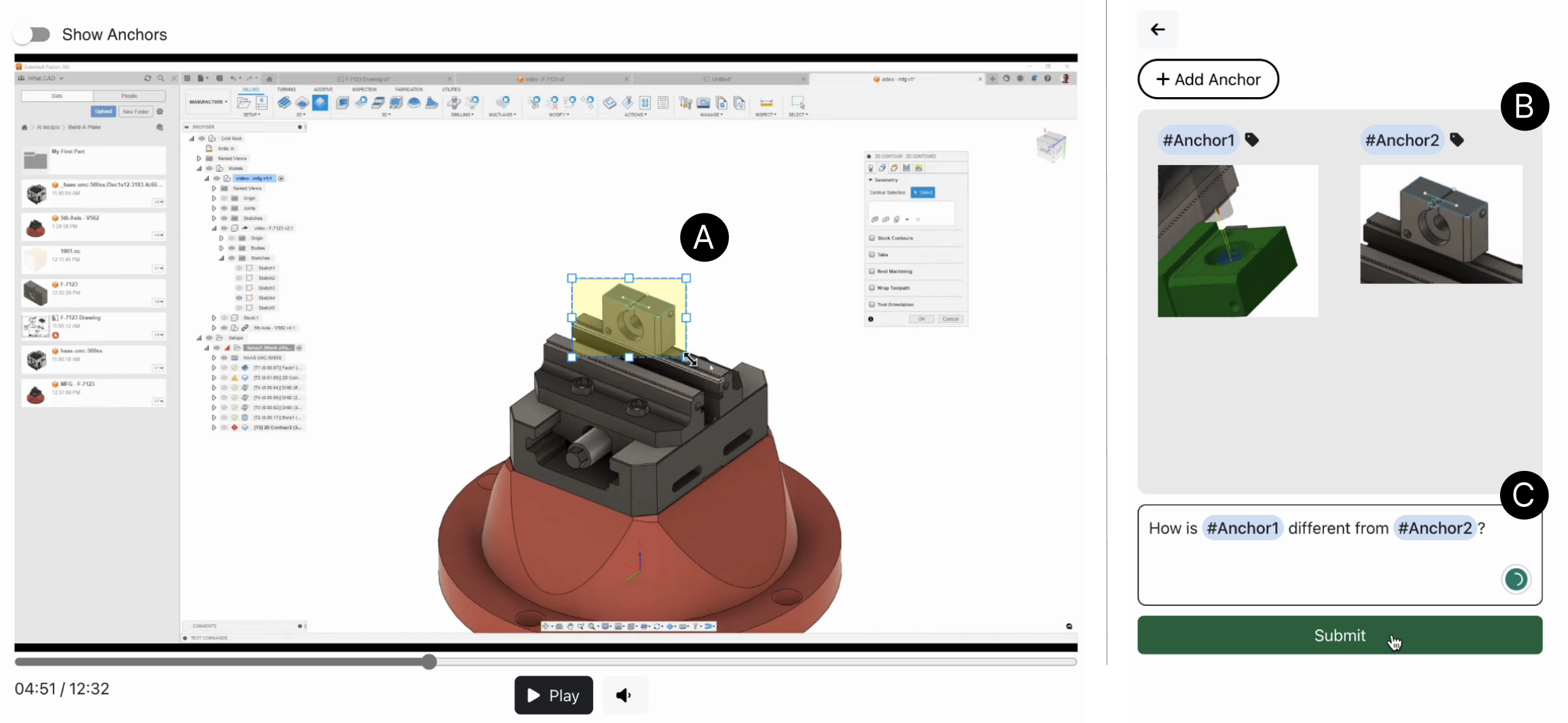}
    \caption{The system used for collecting questions with visual references. (A) Users can draw anchors on parts of the video they want to ask questions about, (B) which will be added to a temporary gallery. (C) Users can refer to each anchor in their questions.}
    \label{fig:system}
    \Description{The system used for collecting questions with visual references. Users can watch the video on the left, where they can draw anchors on parts of the video they want to ask questions about. Confirmed anchors will appear on the right, in a temporary gallery. Users can refer to each anchor in their questions in the text box that is at the bottom.}
% \vspace*{6pt}
\end{figure*}

\subsection{\pointonecolor{Implications on the Answer Pipeline}}
\pointone{In our exploration of automated methods to address questions, we specifically focus on \textbf{`Content'} and \textbf{`User'} questions, as these types of questions are often time-sensitive and crucial for enhancing comprehension and the learning experience with the tutorial content. Moreover, they can often be answered without the involvement of the tutorial authors or software vendors.}

\pointonecolor{
Since these types of questions have direct relevance to video content, a notable trend that emerged from our analysis was the frequent references to video in these questions, sometimes explicitly citing timestamps, which echoes findings from prior research on referencing behavior in comments on a variety of videos~\cite{yarmand2019reference}. It implies that our answer pipeline design should account for what in the video a question was about and the context of the tutorial when the question was posed. Furthermore, since these questions demand a deep understanding of the software, our pipeline should be able to provide accurate and software-specific answers.
}

\pointonecolor{
Additionally, we observed that users often described visual elements of the video in their queries, detailing UI elements within the software (e.g., \textit{``[...] I do not have a `design' option on the left hand drop down and I do not have a `constraints' panel at the top. [...]''}). This behavior of visual references aligns with findings from earlier studies~\cite{yarmand2019reference, snapstream, streamsketch}, which can be particularly prominent in software tutorial videos where the author conveys the workflow via screen sharing~\cite{responsive2015}. However, articulating visual objects in text can be challenging~\cite{kim21winder}, and conventional video interfaces typically support only timestamp references in addition to text. 
This observation prompted our next study, where we aimed to explore the types of references people make when equipped with a tool that allows for visual references.
}

\section{Formative Study 2: Analysis of Questions with Visual References} \label{study2}

% \ar{A notable trend that emerged from our initial analysis was the frequent use of visual references by users when asking questions about the video content. Users referred to specific parts of the video to provide context for their inquiries, detailing the UI elements displayed in the software or the actions carried out by the tutorial's author (e.g., \textit{``[...] I do not have a `design' option on the left hand drop down and I do not have a `constraints' panel at the top. [...]''}). This echos findings from previous research showing that people refer to a visual part of the video to ask questions, leave suggestions, or share their thoughts~\cite{yarmand2019reference, snapstream, streamsketch}. This tendency to rely on visual cues could be especially prominent in software tutorial videos, since the author often conveys the primary workflow via screen sharing~\cite{responsive2015}.
% }

% Building on our results from the first formative study observations, 
\pointfour{To further investigate the visual referencing behavior in software tutorial videos,} we conducted a second study to delve deeper into what people specifically refer to when mentioning specific parts of a video when asking questions. To accomplish this, we ran a data collection study where participants were instructed to watch a software tutorial video, ask questions, and annotate the video to identify visual parts of the video relevant to their questions. Next, we describe the system we built, the study protocol, and the analysis of the results.

\subsection{Data Collection System}
We developed a web application that allows participants to ask questions by directly referring to a specific part of the video (Figure~\ref{fig:system}). Participants can draw an anchor on the video screen (Figure~\ref{fig:system}-A) that they want to ask a question about. These anchors are then saved to a temporary gallery (Figure~\ref{fig:system}-B), which allows participants to directly link to these anchors in their questions (Figure~\ref{fig:system}-C). Anchors can also be labeled with a hashtag (e.g.,\texttt{\#palette}) for easier reference. Clicking on an anchor will populate its label into the question box. Participants can include multiple anchors in a question if they wish to refer to different parts of the video. This can be particularly useful for questions that need to refer to an action that spans a longer segment of the video.

\subsection{Study Design}

\begin{table*}[t]
\centering
\begin{tabular}{c|p{65mm}|c|c|c}
\toprule

{\textbf{Software}}& \hfil \textbf{Video Topic} & \hfil \textbf{Video URL} & \hfil \mr{\textbf{Length}} & \hfil \mr{\textbf{Remark}} \\ \hline

\multirow{4}{*}{Fusion 360} 
& \multicolumn{1}{c|}{Spherical Louvres} & \href{https://youtu.be/K0bKT5PmYx0}{youtu.be/K0bKT5PmYx0} &{5:41} &{TO}\\ \cline{2-5}
& \multicolumn{1}{c|}{4 Ways to Export to DXF} & \href{https://youtu.be/f28TKYsqd6w}{youtu.be/f28TKYsqd6w} &{4:23}&{-}\\ \cline{2-5}
& \multicolumn{1}{c|}{Serrated Washer Using Loft To A Point} & \href{https://youtu.be/fM0AwDLq6_E}{youtu.be/fM0AwDLq6\_E}&{5:31} &{-}\\ \cline{2-5}
& \multicolumn{1}{c|}{Simulating Motion} & \href{https://youtu.be/pGFY-ZXm6G0}{youtu.be/pGFY-ZXm6G0} &{5:22}&{-}\\ \cline{2-5}
\hline

\multirow{4}{*}{Photoshop} 
& \multicolumn{1}{c|}{Poster Design} & \href{https://youtu.be/yFHfOlEVcxs}{youtu.be/yFHfOlEVcxs}&{4:19}&{TO} \\ \cline{2-5}
& \multicolumn{1}{c|}{Soft \& Dreamy Glow Effect} & \href{https://youtu.be/4YaQ5yHQDtg}{youtu.be/4YaQ5yHQDtg}&{4:32}&{-} \\ \cline{2-5}
& \multicolumn{1}{c|}{Using Animation} & \href{https://youtu.be/ugPYmEGiKxs}{youtu.be/ugPYmEGiKxs}&{4:15}&{{PiP, FS, TO}} \\ \cline{2-5}
& \multicolumn{1}{c|}{Applying a Dual Lighting Effect} & \href{https://youtu.be/Q3sa4uraBkk}{youtu.be/Q3sa4uraBkk}&{6:00}&{FS, TO} \\ \cline{2-5}
\hline

\multirow{4}{*}{Excel} 
& \multicolumn{1}{c|}{Creating Pivot Tables} & \href{https://youtu.be/NrUqtE7X05E}{youtu.be/NrUqtE7X05E}&{5:40}&{-} \\ \cline{2-5}
& \multicolumn{1}{c|}{Building a Power Query Function} & \href{https://youtu.be/UOCderIkdXM}{youtu.be/UOCderIkdXM}&{5:01}&{-} \\ \cline{2-5}
& \multicolumn{1}{c|}{Handling Duplicates, Triplicates, Quadruplets} & \href{https://youtu.be/YsC6NYwHanA}{youtu.be/YsC6NYwHanA}&{4:22}&{PiP, FS, TO} \\ \cline{2-5}
& \multicolumn{1}{c|}{CountIf and Pie Charts} & \href{https://youtu.be/8osaUuI-OUU}{youtu.be/8osaUuI-OUU}&{5:58}&{-} \\

\bottomrule
\end{tabular}
\caption{Tutorial videos used in the question collection study. \mr{All videos are screencast tutorial videos. Lengths are in minutes:seconds. \textit{PiP}: The talking head is displayed in picture-in-picture mode. \textit{FS}: The talking head is shown in full screen, occasionally appearing in the video. \textit{TO}: The video includes text overlays.}}
\Description{Table that shows Software, Video Topic, and Video URL used in the question collection study.}
\label{table:videos}
\end{table*}

\subsubsection{Video Selection}
To ensure we cover diverse types of feature-rich software, we selected tutorial videos for three different applications: Autodesk Fusion 360, Adobe Photoshop, and Microsoft Excel. We selected four videos for each application that satisfy the following criteria: (1) between 4 and 6 minutes in length, (2) published within the last three years, and (3) have more than 1,000 views to ensure the quality of the content. The authors manually verified the videos to ensure that the tutorial was high quality, relatively easy to follow, and well explained. This resulted in a total of 12 tutorial videos (Table~\ref{table:videos}). 

\subsubsection{Participants}
We recruited participants from software-specific community forums such as subreddits for these software applications (e.g., r/Fusion360/). We required participants to have at least some prior experience with the software to collect quality questions. To ensure this, they were asked to share details about their level of experience with the software, including how long they have been using it and the main tasks they typically perform with it. We initially recruited 33 participants\pointfour{, of which we excluded 9 after failing quality control measures (see Section~\ref{sec:study2-task})}, resulting in 24 participants who participated in the study (16 male, 6 female, 2 non-binary, mean age=30.2). The 24 participants were divided evenly over each software application, with 8 participants for each. Each of the 12 selected tutorial videos was assigned to two participants. \pointfour{As the study was done online, participants were expected to be able to access the web with their own desktop or laptop.}

\subsubsection{Task}\label{sec:study2-task}
\pointfour{The study was conducted in an asynchronous remote setting.} After the researchers confirmed a participant's eligibility regarding their experience with the software, the participant received a URL to our system. The process began with an informed consent form, followed by a demographic survey. Participants then went through a brief tutorial detailing how to use our system to ask questions and make visual references. \pointfour{They were instructed to imagine themselves as someone watching this video to improve their skills and ask questions about the tutorial. Specifically,} participants were asked to pose at least 10 questions, each accompanied by one or more visual anchors, which are visual elements in the video the question is about. They could redraw an anchor until they were satisfied with it. They could also rename the anchors and easily refer to them in the question by either clicking on the anchor or typing its name. At the end of the study, we asked for optional open-ended feedback about the overall study.
\pointfour{As a quality control measure, we excluded participants with more than five non-question responses. These included those expressing appreciation for the tutorial video (e.g., "\textit{This is well explained}") or making suggestions (e.g., "\textit{It would be better to specify a number here.}")}
% \mr{As a quality control measure, we excluded participants for whom more than half of their questions were not questions.} 
% "Please add subtitles."
The study took around 40 minutes, and the participants were compensated with a \$30 USD gift card for their participation.

\subsection{Results}\label{sec:study2-results}
% 217 questions, 237 anchors, 2 anchors: 16, 3 anchors: 2
From the study, we collected a total of 256 responses \pointfour{from 24 participants}, each accompanied by at least one relevant visual anchor. 
We filtered out non-question responses, such as those expressing appreciation or making suggestions, leaving us with 217 questions in total. 
\pointfour{These questions were composed of 205 explicit questions and 12 questions which implicitly sought help (e.g., "\textit{\#Anchor1 doesn't seem to work and I have no idea how to deal with it.}")}.
Below, we present the analysis of questions and associated visual anchors that we collected.

\subsubsection{Questions and Visual Anchors}
Each question was accompanied by one or more visual anchors. While most of the questions (91.7\%) were associated with a single visual anchor, 18 out of 217 questions had two or more visual anchors for elaboration. Specifically, 16 questions used 2 anchors, and 2 questions used 3. These multiple anchors served various purposes, such as indicating the beginning and end of an action, posing questions about a `before' and `after' scenario, and suggesting alternative options to consider.

\begin{table}[t]
\centering
\begin{tabular}{>{\centering\arraybackslash}m{1.8cm}p{6cm}}
\toprule
\textbf{Type} & \textbf{Visual Anchor \& Question} \\
\midrule
UI Elements & \parbox[c]{6cm}{\vspace{2pt}\raisebox{-.5\height}{\includegraphics[width=0.08\textwidth]{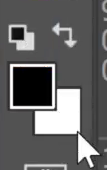}}\vspace{2pt} \\ Can you explain more about what that icon is used for?} \\
\midrule
Workspace & \parbox[c]{6cm}{\vspace{2pt}\raisebox{-.5\height}{\includegraphics[width=0.3\textwidth]{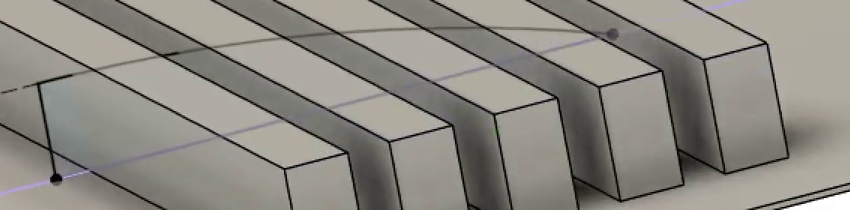}}\vspace{2pt} \\ How do I determine my pattern spacing?} \\
\midrule
UI+Workspace & \parbox[c]{6cm}{\vspace{2pt}\raisebox{-.5\height}{\includegraphics[width=0.3\textwidth]{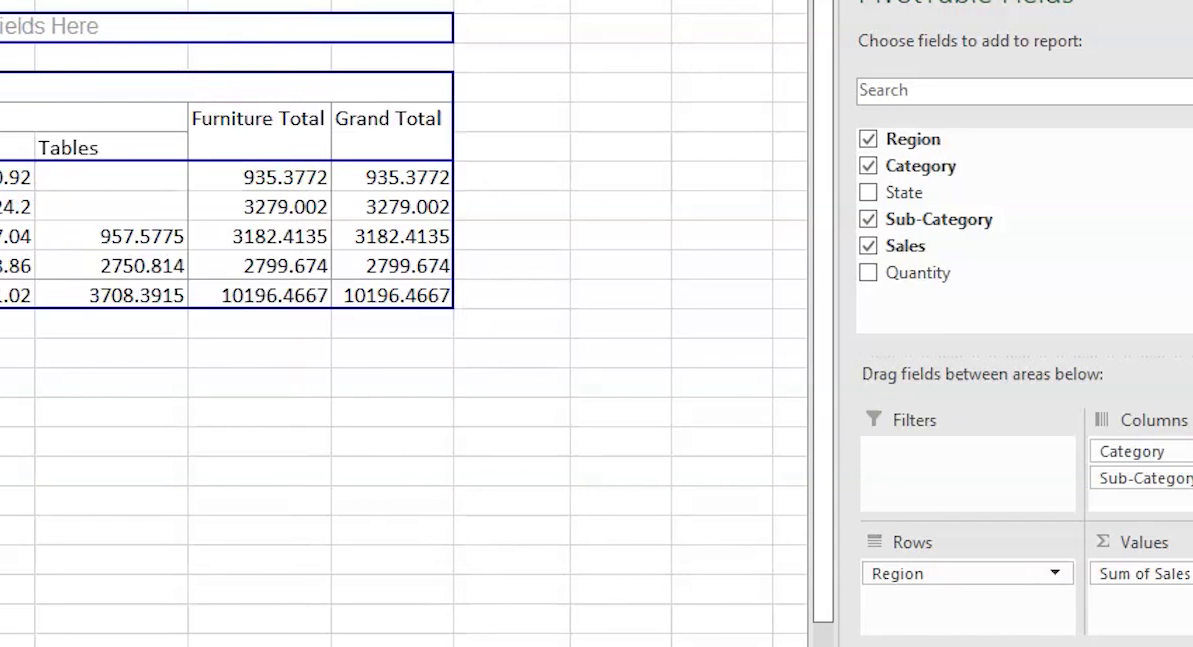}}\vspace{2pt} \\ Can you put a second field into the Row section as well as the Columns section?} \\
\midrule
Annotation & \parbox[c]{6cm}{\vspace{2pt}\raisebox{-.5\height}{\includegraphics[width=0.3\textwidth]{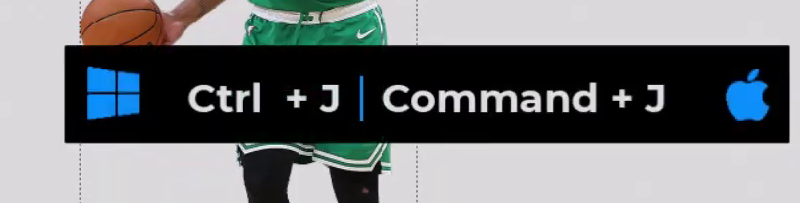}}\vspace{2pt} \\ What is the alternative command for this?} \\
\midrule
Miscellaneous & \parbox[c]{6cm}{\vspace{2pt}\raisebox{-.5\height}{\includegraphics[width=0.15\textwidth]{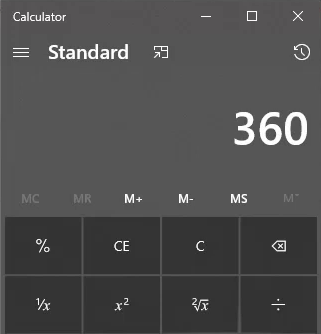}}\vspace{2pt} \\ Where did the calculator come from? Can you do this inside Fusion? How?} \\
\bottomrule

\end{tabular}
\caption{Types of visual anchors and associated questions collected in our study.}
\Description{Table that shows Type, Visual Anchor, and Question collected in our study. Visual anchor in each row represents a UI element in Photoshop, Workspace in Fusion 360, UI and workspace in Excel, Textual annotation on a Photoshop video, and an external calculator application, respectively.}
\label{table:anchor_type}
\end{table}

\subsubsection{Visual Anchor Types}
We first analyzed the types of visual anchors that the participants used in their questions. 
\pointfour{Two authors initially discussed the types of visual anchors based on their roles in the software. Following this, the lead author annotated each visual anchor according to its type and no ambiguous cases were encountered.
Specifically,} our analysis identified five distinct types of visual anchors:

% tool (18.4\%), panel (63.2\%), menu (13.6\%), and input field (4.8\%).
\begin{itemize}
    \item \textbf{UI Elements (52.7\%):} Anchors on User Interface elements like tools, menus, and panels within the software.
    \item \textbf{Workspace (34.6\%):} Anchors placed on the workspace of the software where the tutorial creator is performing the main task, such as the 3D CAD model in Fusion 360, the photo in Photoshop, and the grid of cells in Excel.
    \item \textbf{UI+Workspace (5.5\%):} Anchors that capture both UI Elements and Workspace, including a full screenshot of the interface.
    \item \textbf{Annotation (3.8\%):} Anchors attached to textual or graphical annotations that the tutorial creator has overlaid on the primary video footage.
    \item \textbf{Miscellaneous (3.4\%):} Anchors that are not related to the software, such as those placed on external applications, video annotations (like pop-ups showing entered keyboard commands), or the face of the tutorial creator.
\end{itemize}

Most anchors were related to either the UI elements (52.7\%) or the workspace (34.6\%), with some capturing both (5.5\%). Combining these three categories amounted to 92.8\% of anchors. The questions associated with these anchors involved asking about the functionality of specific tools or about detailed methods of the workflow. Table~\ref{table:anchor_type} shows example questions and related visual anchors for each type.

%%%%%%% EXAMPLE TABLE %%%%%%%%

\subsubsection{Role of Visual Anchors}

%%%%%%% EXAMPLE TABLE (ROLE) %%%%%%%%

\begin{table}[t]
\centering
\begin{tabular}{>{\centering\arraybackslash}m{1.8cm}p{6cm}}
\toprule
\textbf{Role} & \textbf{Visual Anchor \& Question} \\ 
\midrule
Necessary
& \parbox[c]{6cm}{\vspace{2pt}\raisebox{-.5\height}{\includegraphics[width=0.1\textwidth]{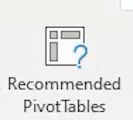}}\vspace{2pt}} \\
& Does \textbf{this feature} provide the same result as the pivot table template chosen in this video? \\
\midrule
Useful 
& \parbox[c]{6cm}{\vspace{2pt}\raisebox{-.5\height}{\includegraphics[width=0.2\textwidth]{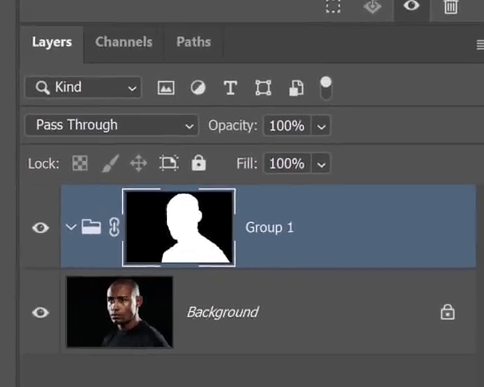}}\vspace{2pt}} \\
& Can you explain more about \textbf{the layer mask option}? What is it used for? and its uses? \\
\midrule
Irrelevant 
& \parbox[c]{6cm}{\vspace{2pt}\raisebox{-.5\height}{\includegraphics[width=0.18\textwidth]{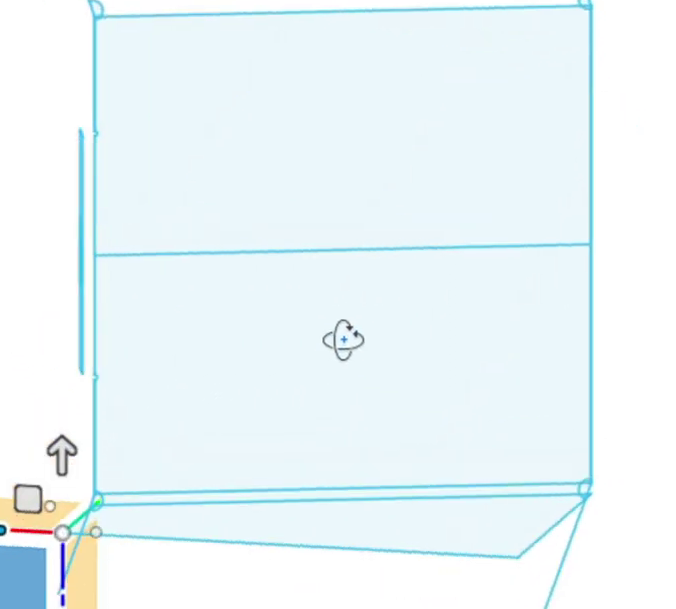}}\vspace{2pt}} \\
& What kind of file does it create that can't be opened by these programs? \\
\bottomrule

\end{tabular}
\caption{Roles of visual anchors and associated questions collected in our study. Text in bold refers to the associated visual anchor.}
\Description{Table that shows Role, Visual Anchor, and Question collected in our study. Visual anchor in each row represents a UI element in Excel, a UI element in Photoshop, and a workspace in Fusion 360, respectively.}
\label{table:anchor_role}
\end{table}

We then examined the level of involvement of visual anchors in the questions. 
\pointfour{We first found that the visual anchors could either be crucial for interpreting the questions (\textit{Necessary}), merely add extra context (\textit{\mr{Useful}}), or not relevant to the question at all (\textit{Irrelevant}): 
% We sought to understand whether the visual anchors were crucial for interpreting the questions (\textit{Necessary}), merely added extra context (\textit{\mr{Useful}}), or were irrelevant (\textit{irrelevant}):
\begin{itemize}
    \item \textbf{Necessary (47.5\%):} The question by itself is unclear or lacks context, and thereby visual anchors are required to fully comprehend the question.
    \item \textbf{\mr{Useful} (49.3\%):} The question is understandable without additional context, and visual anchors are relevant to the question’s context.
    \item \textbf{Irrelevant (3.2\%):} Visual anchors have no connection to the question.
\end{itemize}
The lead author annotated each visual anchor according to its role. Ambiguous cases (16 out of 235) were resolved through discussion with another author. 
}

We identified that almost half of the questions (47.5\%) required the accompanying visual anchors to be fully understood. These questions often used referential terms like ``this'' or ``it'', which directly pointed to the visual anchors for context. The remaining half (49.3\%) mostly consisted of questions where the anchors played a supportive role. Only a small fraction of questions (3.2\%) were found to have irrelevant anchors. These questions were mostly associated with what the tutorial author talked about when the visual anchor was drawn, or they were broad questions related to the general topic of the tutorial. Table~\ref{table:anchor_role} provides examples of questions and their corresponding visual anchors, categorized by the role each anchor played.

% \begin{figure}
%     \centering
%     \begin{subfigure}{0.45\textwidth}
%         \includegraphics[width=\linewidth]{figure/chart1.pdf}
%         \caption{Number of visual anchors for each type}
%     \end{subfigure}
%     \hfill
%     \begin{subfigure}{0.45\textwidth}
%         \includegraphics[width=\linewidth]{figure/chart2.pdf}
%         \caption{Distribution of role of visual Anchors}
%     \end{subfigure}
%     \caption{Analysis results of visual anchors made in questions.}
% \end{figure}

\subsection{\pointonecolor{Implications on the Answer Pipeline}}
\pointone{
From the study, we could see that people mostly refer to software UI elements or the workspace when asking questions, and nearly half of the questions required these visual anchors to provide important contextual information. These findings highlight the importance of comprehending visual references associated with a question, particularly those related to software UI elements, which were found to be the majority of visual references.
}

\subsection{Design Goals}\label{sec:design}
\pointone{
From the two formative studies, we derived the following design goals of an answer pipeline for software tutorial videos. First, our answer pipeline should consider what was happening in the video when the question was asked (DG1). Second, it should be able to understand (multiple) visual references associated with a question, including software UI elements (DG2). Lastly, it should provide accurate and useful answers that reflect the software-specific knowledge (DG3).
\begin{itemize}
    \item DG1: Consider the video context when the question was posed.
    \item DG2: Understand visual references associated with the question, including software UI elements.
    \item DG3: Provide accurate and useful software-specific information in answers.
    % Support multiple visual references?
\end{itemize}
}
\section{{\emph{AQuA}}: Question-Answer Pipeline}\label{sec:pipeline}
Based on our \pointone{design goals (Section~\ref{sec:design})}, we designed \emph{AQuA}, a question-answer pipeline that generates useful responses to questions with visual anchors (Figure~\ref{fig:pipeline}). 
%The pipeline has the following design objectives: \pointone{(DG1) The ability to understand the tutorial context at the moment the question was asked, (DG2) the ability to interpret and understand visual anchors in regard to the software's functionality, and (DG3) the ability to offer specific and accurate responses to the questions relating to the software's functionality and workflows.} 
We applied our pipeline to Fusion 360 to demonstrate the potential of this approach. While we tailored the pipeline to have specialized knowledge about Fusion 360, this approach is generalizable to other feature-rich software applications. \emph{AQuA} will work for any software application for which tool or command names accompanied by icons or screenshots of corresponding UI elements, and a sufficiently large set of software documentation or existing tutorial materials are available. This would, for example, be the case for many other feature-rich software applications such as Adobe Photoshop or Illustrator, Autodesk AutoCAD or Maya, or Microsoft Word.

\subsection{Overall Architecture of \emph{AQuA}}

\begin{figure*}[t]
    \centering
    \includegraphics[width=0.9\linewidth]{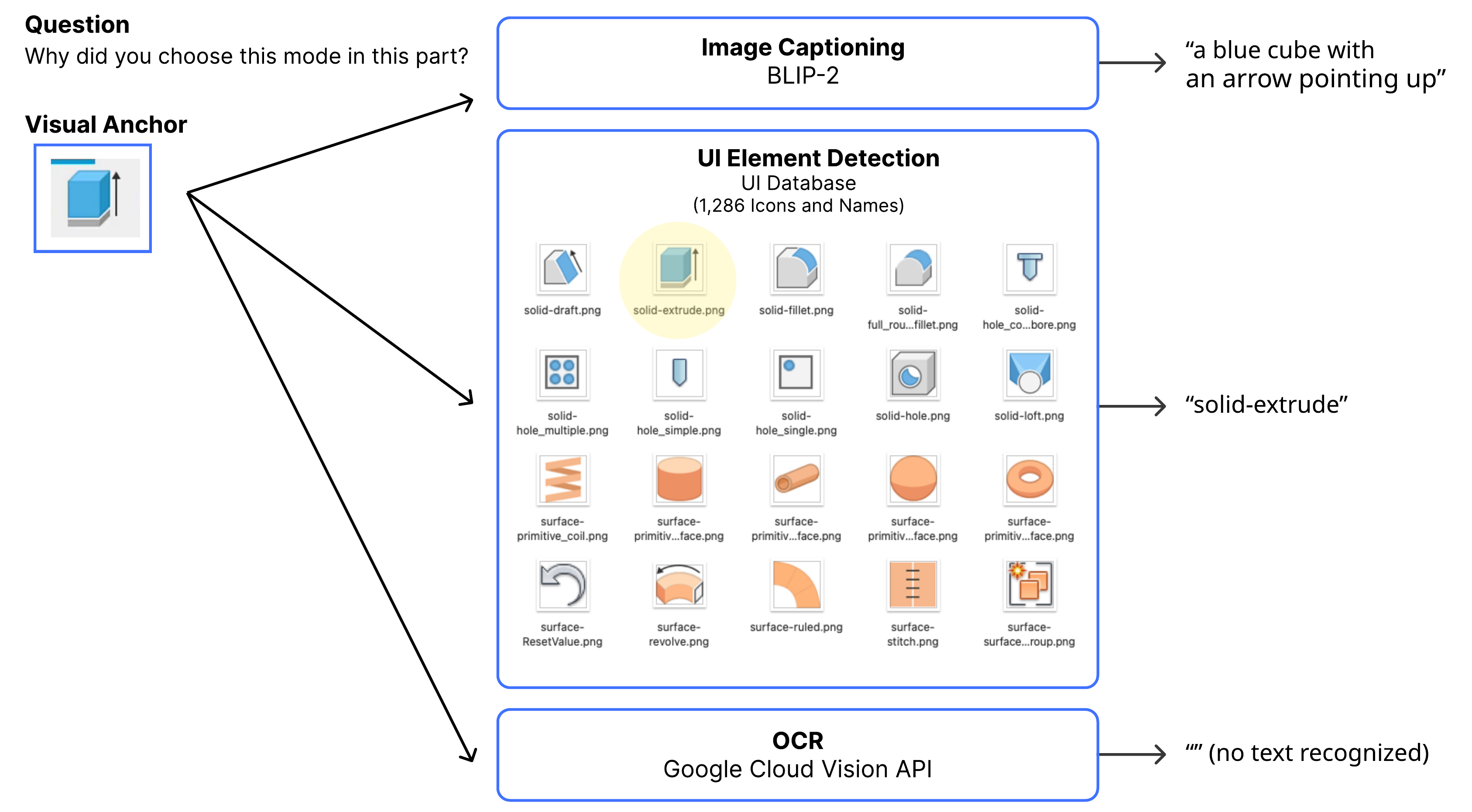}
    \caption{\pointfive{Our Visual Recognition Module is composed of Image Captioning, UI Element Detection, and Optical Character Recognition (OCR). We use BLIP-2~\cite{li2023blip2} to obtain a general description of the visual anchor in case it contains generic or workspace objects, and the Google Cloud Vision API~\cite{cloud-vision-api} to detect any textual information in the anchor. For UI Element Detection, we first run UIED~\cite{uied} to determine if there are multiple UI elements in the anchor. Then, we apply feature matching and template matching between each element in the anchor and those in the UI database. If the matching score exceeds a certain threshold, we retrieve the element's name.}}
    \label{fig:ui_database}
    \Description{An example of how our visual recognition module works. It shows a question and a visual anchor and Image Captioning, UI Element Detection, and OCR components.}
% \vspace*{6pt}
\end{figure*}

Figure~\ref{fig:pipeline} illustrates the overall architecture of our question-answer pipeline. It takes the question text and visual anchor(s) as inputs. First, our Visual Recognition Module identifies the UI element in the visual anchor and generates a textual description of it (\pointone{DG2, }Section~\ref{sec:pipeline_visual}) so that it can be easily provided to GPT-4 (which, at the time of writing, did not yet support multimodal prompts in its API). Then, we combine the visual description with the question text and search a database of software documentation and tutorial materials for articles relevant to the query (\pointone{DG3, }Section~\ref{sec:pipeline_retrieval}). Along with these retrieved articles, we include the video title and relevant transcript sentences to give context to the question (\pointone{DG1, }Section~\ref{sec:pipeline_video}).
All of these elements---question text, visual anchor descriptions, retrieved articles, and video context---are fed into GPT-4 through crafted prompts. These prompts instruct the model to provide answers to questions by specifying the components in the following order:
Relevant articles; tutorial titles and transcripts; questions; and visual anchors. 
The prompts provided as input for GPT-4 can be found in Appendix~\ref{sec:prompts_full}.

% \begin{lstlisting}[caption={Prompts Used in Our Question-Answer Pipeline},label=lst:full_prompt]
% You need to answer questions about Autodesk Fusion 360 that people asked while watching a tutorial video. Please answer in 50 words or less. Each question is accompanied by relevant visual anchors, which are specific visual elements of interest in the video.

% Use the below articles on the Fusion 360 software to answer the subsequent question. If the answer cannot be found in the articles, write ``I could not find an answer.''
% Fusion 360 article section: {section 1}
% Fusion 360 article section: {section 2}
% ...

% Tutorial: Title: {title}. Instructions: {transcript}
% Question: {question_text}
% Visual Anchor: 
% {Anchor_label_1}: {blip}. It includes the Fusion 360 tools: {tool} and text: {ocr}.
% {Anchor_label_2}: {blip}. It includes the Fusion 360 tools: {tool} and text: {ocr}.
% ...

% \end{lstlisting}

\subsection{Visual Recognition Module}\label{sec:pipeline_visual}

To identify visual anchors and generate textual descriptions, our pipeline includes a Visual Recognition Module (Figure~\ref{fig:ui_database}) that is composed of Image Captioning, UI Element Detection, and Optical Character Recognition (OCR). Below, we explain these submodules.

\subsubsection{Image Captioning}
To obtain descriptions of general visual anchors such as parts of the application workspace, we use BLIP-2~\cite{li2023blip2}, a visual-language model that shows high performance on zero-shot image captioning. BLIP-2 recognizes objects within an image, thereby constructing a well-defined image description that captures the core information. This can be particularly useful when the visual anchor contains objects that the tutorial author is working on (e.g., \textit{``a gray steel washer on a white background'') or generic objects that the author pulls up (e.g., \textit{``a calculator with the number 360 on it''}).}

\subsubsection{UI Element Detection}
% 447: documentation 881: command
Our formative study revealed that \nr{more than half} of the visual anchors are related to UI elements. While visual-language models like BLIP-2 are excellent at describing real-life scenes (e.g., \textit{``a couple with a dog on a leash on the beach at sunset''}) \pointfive{or general appearances of objects (e.g., \textit{``a cylindrical object with a hole in it''})}, they fall short in recognizing software-specific elements. For instance, for the visual anchor shown in \mr{Figure~\ref{fig:ui_database}}, BLIP-2 generates the description: \texttt{a blue cube with an arrow pointing up}, which, while not entirely wrong, is too generic and not helpful to identify the icon in the visual anchor as the \texttt{Extrude} tool in Fusion 360.

To more accurately recognize these UI elements, we created a UI element image database. This was done by crawling software help resources from the official Fusion~360 documentation~\cite{fusion-documentation}. We exploited the fact that the tool image and its name follow a specific pattern arranged in HTML unordered list items, with the tool name preceding the image. To extract the name and image, we accordingly parsed the HTML files based on this identified pattern. To further enrich our data, we ran Fusion~360 and extracted additional UI information by running a script to save all command icons. This approach yielded a total of 1,286 images along with their corresponding names---446 from the documentation and 840 from the software commands.

By leveraging this UI icon database that we created, our pipeline is able to identify UI elements within visual anchors by comparing them to the images in the database. If the dimensions of an image exceed a certain threshold (in our case, 100 pixels in both width and height), we first proceed with an initial UI element detection pass, as the image could contain multiple elements due to its size. We use UIED~\cite{uied} to generate bounding boxes around each UI element within the image. Following this, we apply an image similarity algorithm to each bounding box to identify the corresponding image in our database. First, we use OpenCV's feature matching~\cite{feature-matching} to extract and compare visual features, thereby establishing matched features. We then select the top five candidate images based on the number of matched features. To refine our search, we apply template matching to the five candidate images using the Normalized Cross-Correlation Coefficient~\cite{template-matching} to locate instances of a template image within a larger search image. If the highest template score exceeds 0.5, we deem it a successful match and retrieve the element's name.

\subsubsection{OCR}
Lastly, since UI elements often contain text~\cite{sikuli}, we employ Optical Character Recognition (OCR) to extract textual information using the Google Cloud Vision API~\cite{cloud-vision-api}. This can be particularly useful when the visual anchor includes a menu or panel that lists the names of various functionalities. 

We run the above three modules and combine their output to generate a final textual description for each visual anchor associated with a question. \pointfive{This description is provided as part of a prompt to GPT-4 when generating answers and is also used to retrieve relevant materials in the Retrieval Module (Section~\ref{sec:pipeline_retrieval}).}

\begin{table*}[t]
\centering
\begin{tabular}{c|c|c|c|c}
\toprule
Batch & Participant & Occupation & \makecell{Experience \\ (years)} & Main Tasks \\
\hline
\multirow{8}{*}{Batch 1} 
& P1 & CAD Engineer/Project Manager & 4-5 & Assemblies, Sheet Metal \\
& P2 & 3D Printer Technician & 4 & 3D Modeling, Assemblies \\
& P3 & Freelancer & 4 & Simulation, 3D Modeling, Data Management \\
& P4 & Software Developer/Designer & 4 & 3D Graphic Design \\
& P5 & Retired & 7 & 3D Modeling \\
& P6 & Freelancer & 1 & Simulation, Data Management \\
& P7 & Data Analyst & 10 & Simulation \\
& P8 & Localization Program Manager & 1 & Design \\
\cline{1-5}
\multirow{8}{*}{Batch 2} 
& P9 & Consultant & 7 & 3D Modeling \\
& P10 & Software Developer & 3 & 3D Graphic Design, Simulation \\
& P11 & Mechanical Design Engineer & 5 & 3D Modeling, Rendering, Manufacturing \\
& P12 & Customer Service Representative & 8 & Parametric Modeling, Drawing, Renders \\
& P13 & Routesetter & 2-3 & 3D Modeling \\
& P14 & Product Designer & 4 & 3D Modeling, Product Design \\
& P15 & Product Manager & 8 & Manufacturing, Design \\
& P16 & Content Strategist & 3 & Demonstration \\

\bottomrule
\end{tabular}
\caption{Participant information including current occupation, number of years of experience, and the main tasks they perform with Fusion 360.}
\Description{Table that shows Batch, participant number, occupation, number of years of experience, and the main tasks they perform with Fusion 360.}
\label{table:participants}
\end{table*}

\subsection{Retrieval Module}\label{sec:pipeline_retrieval}
Large Language Models such as GPT-4 can suffer from hallucinations, which is when they generate false information~\cite{zhang2023hallucination}. To enhance the quality of responses and make GPT-4 generate answers specific to the software, we further enrich the pipeline \pointfive{with Retrieval Augmented Generation (RAG). RAG is an approach that combines retrieval and generation, producing specific and factual responses~\cite{rag20}. This is achieved by incorporating relevant information retrieved from a knowledge base when generating a response to a prompt.
In our case, we constructed a knowledge base by gathering Fusion 360 articles.}
We used the same documentation source used in Section~\ref{sec:pipeline_visual}, which contains 2,937 HTML files. Additionally, we gathered 2,375 Fusion 360 tutorial videos from Autodesk Screencast~\cite{screencast, chronicle} and transcribed their audio using Amazon Transcribe~\cite{amazon}. We segmented each content into chunks, ensuring that each chunk did not exceed a certain length (i.e., 1,600 tokens). We then obtained embeddings for each chunk using OpenAI's text embedding model (\texttt{text-embedding-ada-002})~\cite{openai-embedding}. This resulted in a total of 5,635 \pointfive{\emph{article chunks} (i.e., either documentation or tutorial transcripts)} and their embeddings. 

Using the same embedding model, our pipeline gets embeddings of the provided question text and visual anchor description. \pointfive{We then compare these embeddings with that of each article in the knowledge base we constructed, retrieving the top 50 articles based on the cosine similarity ranking between the embeddings.} The top 50 articles represent the top 1\% among all articles, but the number of articles used by the pipeline typically ranges around the top 20, depending on their length. This is because we stop appending articles to the collection once adding another exceeds the input token limit of GPT-4 (i.e., 8,192 tokens in our case). The resulting collection of articles is then provided as part of a prompt to GPT-4 when generating answers.

\subsection{Video Context }\label{sec:pipeline_video}
Finally, to provide the video context \mr{when the question was posed}, we include both the title and a relevant segment of the transcript of the tutorial video about which the question was asked. We extract the two transcript sentences that are adjacent to the timestamp where the visual anchor was captured. Specifically, we locate the nearest sentence whose starting timestamp does not exceed the timestamp of \pointfive{when the visual anchor was captured}, and then extract both that sentence and the preceding one. If there are multiple visual anchors, the two relevant transcript sentences for each visual anchor are concatenated. 
\pointfive{These title and relevant transcript sentences are provided as part of a prompt to GPT-4 when generating answers.}
% Transcribed using Whisper

For demonstration and testing purposes, we transcribed the 12 videos we used in the second formative study (Section~\ref{study2}) using Whisper~\cite{radford2022whisper}. However, given that popular websites for tutorial videos, such as YouTube, already support downloading transcripts for many videos and continue to improve in accuracy, we imagine that in the future, transcripts with detailed timestamps could be easily provided as input to the pipeline in addition to the video.

\section{Evaluation}\label{sec:evaluation}

\begin{figure*}[t]
    \centering
    \includegraphics[width=\linewidth]{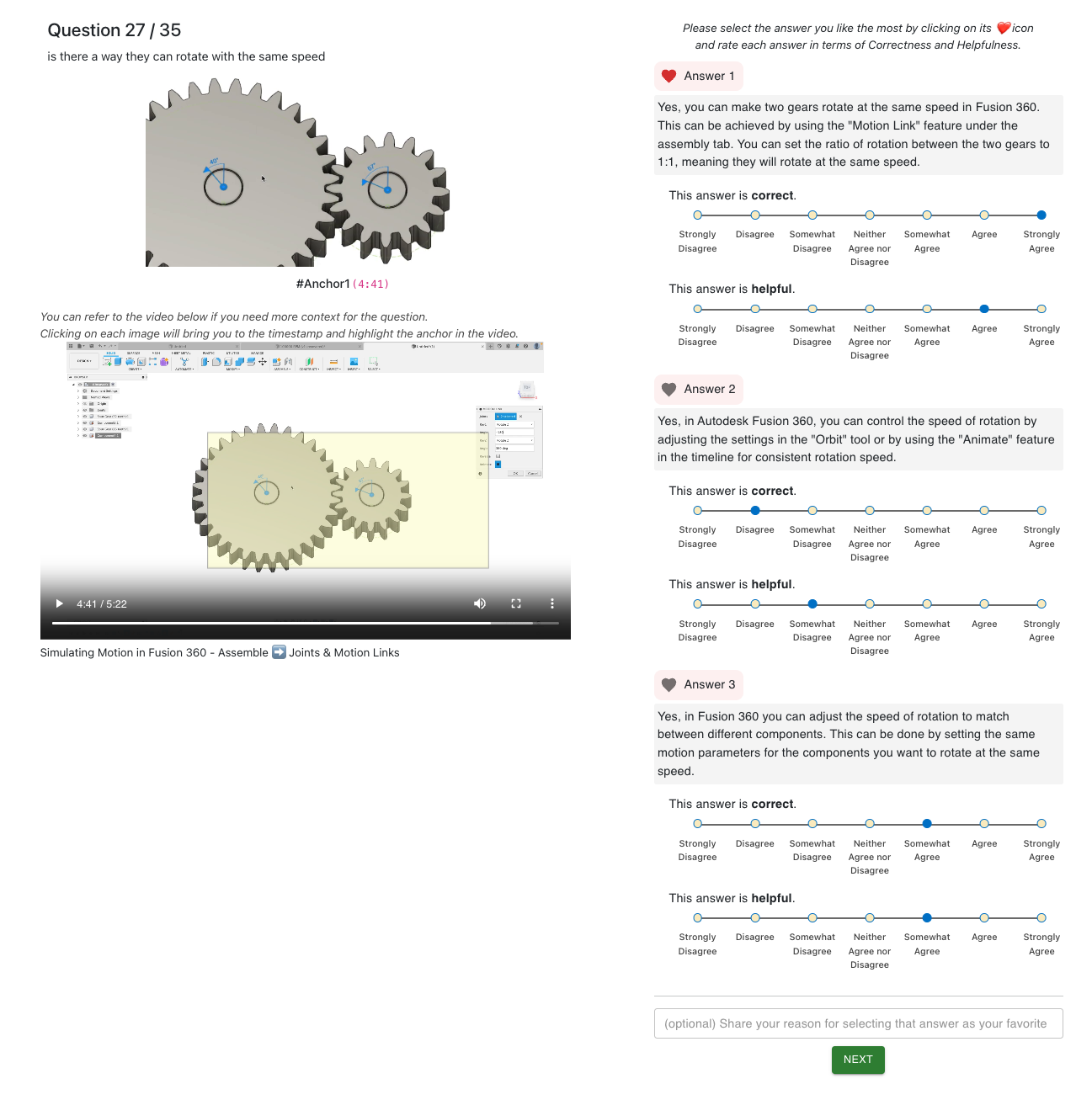}
    \caption{The system used in our pipeline evaluation study. The participant can see the question, the video that the question was asked about at the right timestamp and with the visual anchor highlighted, and three generated answers in random order. They were asked to rate each answer in terms of its correctness and helpfulness on a scale of 1 to 7, and select their favorite answer among the three. Optionally, they could provide reasons for selecting their favorite answer.}
    \label{fig:eval_system}
    \Description{The system used in our pipeline evaluation study. On the left side, the participant can see the question at the top and below the video that the question was asked about at the right timestamp and with the visual anchor highlighted. On the right side, three generated answers are presented in random order. Each answer is accompanied by two 7-point Likert scales asking about its correctness and helpfulness. Participants were asked to select their favorite answer among the three by clicking a heart icon next to the answer. An optional text box where they could provide reasons for selecting their favorite answer is presented at the bottom.}
% \vspace*{6pt}
\end{figure*}

We evaluated our proposed question-answer pipeline with 69 questions from four Fusion 360 videos that were gathered during our formative study (Section~\ref{study2}). \sr{These questions consisted of 87\% `Content' questions and 13\% `User' questions.} We generated answers to these questions under three different conditions: \textbf{(1) \texttt{Question-only}}, where only the question text is provided to GPT-4; \textbf{(2) \texttt{Question + Video}}, where the question text together with the video title and transcript sentences are provided to GPT-4; and \textbf{(3) \texttt{Full Pipeline (AQuA)}}, in which also the visual anchor and relevant articles are provided to GPT-4. We treat \texttt{Question-only} and \texttt{Question + Video} as baseline methods, since the former could be seen as a plain GPT-4 while the latter could be considered a plain GPT-4 with a bit more information from the video. We instructed GPT-4 to generate answers in 50 words or less in all conditions. To reduce variability in GPT-4's responses, we set the temperature parameter to 0, which minimizes randomness in the generated answers. We also used GPT-4-0613, a snapshot of GPT-4 that was available at the time of our study. The prompts used for each condition can be found in Appendix~\ref{sec:prompts}.

\subsection{Participants}
We recruited participants with varying levels of experience in Fusion 360 (Table~\ref{table:participants}) to evaluate the answers generated under each condition. We initially recruited 21 participants, of which we excluded 5 after failing quality control checks (see Section~\ref{sec:eval-task-and-measures}), resulting in 16 (2 female, 14 male, mean age=34.5) who participated in the study. We recruited participants from a Fusion 360 subreddit (N=12) as well as people who develop the Fusion~360 software as part of their day job (N=4). We divided the 69 questions into two batches to accommodate for a reasonable study duration (30--60 minutes). The two batches contained 34 and 35 questions respectively, and we assigned 8 participants to each batch. 
%While this meant not all questions were evaluated by the same set of users, we decided this trade-off was worth it to ensure we could evaluate the largest set of 69 questions to obtain a clear picture of the performance of each answer condition and of our \emph{AQuA} pipeline. 
Where appropriate, we discuss the results for each batch separately.

\begin{table*}[h]
\centering
\begin{tabular}{p{38mm}|p{38mm}|p{38mm}|p{38mm}}
\toprule
Question \& Visual Anchor & \texttt{Question-only} & \texttt{Question + Video} & \texttt{Full Pipeline (AQuA)} \\
\hline
(1) How did you get this menu to appear? \includegraphics[width=0.15\textwidth]{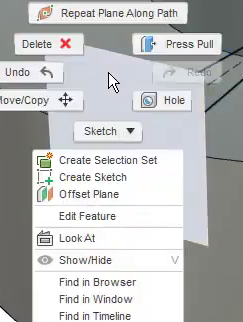} & The menu in Autodesk Fusion 360 typically appears by either clicking on the relevant icon on the toolbar or by right-clicking to bring up a context menu. \textbf{The exact method may vary depending on the specific menu you're referring to in the tutorial.} & The menu in Autodesk Fusion 360 typically appears when you right-click on the workspace or when you select a specific tool from the toolbar at the top of the screen. \textbf{In this case, the menu likely appeared when the instructor selected the "Create Sketch" tool}. & \textbf{The menu shown is the Marking Menu in Autodesk Fusion 360}. It appears when you right-click anywhere in the canvas. This radial menu provides quick access to frequently used commands. To activate a command, drag the cursor in the direction of the command, then click anywhere in the highlighted wedge. \\ \hline

(2) Is there a way they can rotate with the same speed? 
\includegraphics[width=0.2\textwidth]{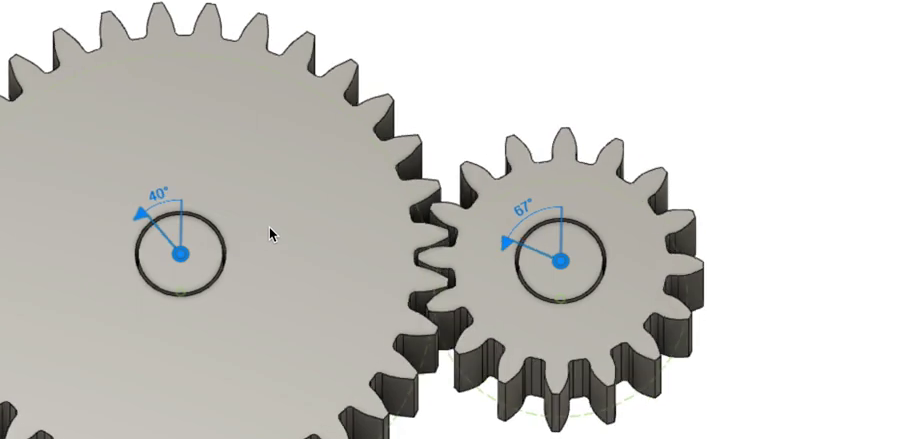} & Yes, in Autodesk Fusion 360, you can control the speed of rotation by adjusting the settings in the \textbf{"Orbit" tool or by using the "Animate" feature} in the timeline for consistent rotation speed. & Yes, in Fusion 360 you can adjust the speed of rotation to match between different components. \textbf{This can be done by setting the same motion parameters} for the components you want to rotate at the same speed. & Yes, you can make two gears rotate at the same speed in Fusion 360. \textbf{This can be achieved by using the "Motion Link" feature} under the assembly tab. You can \textbf{set the ratio of rotation between the two gears to 1:1}, meaning they will rotate at the same speed. \\ \hline

(3) Why are we combining the bodies? \includegraphics[width=0.2\textwidth]{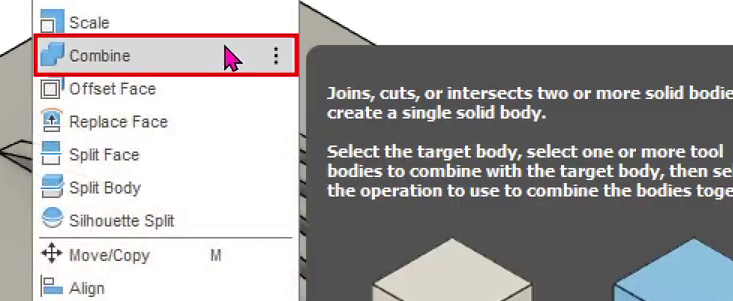} & Combining bodies in Fusion 360 \textbf{helps to create a single, unified object}. This is useful for simplifying the design process, reducing complexity, and ensuring that different parts fit together correctly in the final product. & Combining the bodies in Fusion 360 allows you to \textbf{merge multiple bodies into a single body}. This is useful for creating complex shapes and structures, and for simplifying your design process. & Combining bodies in Fusion 360 allows you to \textbf{join, cut, or intersect multiple solid bodies to create a single solid body.} This is useful for creating complex shapes or structures that are composed of multiple parts. It can also be used to cut out sections of a body using another body as a tool. \\ 

\bottomrule
\end{tabular}
\caption{Example answers generated from each condition. (1) \texttt{Full Pipeline (AQuA)} accurately identifies the exact menu name and provides relevant information, whereas \texttt{Question + Video} refers incorrectly to a menu by relying on the transcript. (2) \texttt{Full Pipeline} provides detailed instructions while \texttt{Question-only} generates inaccurate information. (3) \texttt{Full Pipeline} offers additional details about various operations, making it more informative than others. Note that typos in the questions have been corrected for clarity; however, they were not corrected when generating answers.}
\label{table:answer_examples}
\Description{Table that shows Question and Visual Anchor, answers generated from Question-only, Question + Video, and Full Pipeline (AQuA). Visual anchor in each row represents a UI element, Workspace, and another UI element in Fusion 360, respectively.}
\end{table*}
% 35, 43, 61

\subsection{Task and Measures}\label{sec:eval-task-and-measures}
Participants were presented with each question, along with related visual anchors and the tutorial video that highlights the part where the anchor was captured (Figure~\ref{fig:eval_system}). The answers generated under the three conditions were randomly ordered for each question. Following the metrics of free-form question-answering in natural language generation~\cite{metric}, participants were asked to rate each answer in terms of its correctness (i.e., how accurate the answer is) and helpfulness (i.e., how well the answer addresses the question) on a 7-point Likert scale. Additionally, they were asked to choose their preferred answer among the three options and could optionally provide the rationale behind their choice. As a quality control measure, we excluded participants who had a preferred answer ranked lower than another option on both correctness and helpfulness. This resulted in 2 and 3 participants being excluded from Batch 1 and 2, respectively. At the end of the study, we asked for optional open-ended feedback about the overall answers and the study. For their participation in the 60-minute study, participants were compensated with a \$60 USD gift card.

\subsection{Results}

\begin{figure*}[t]
    \centering
    \includegraphics[width=0.8\linewidth]{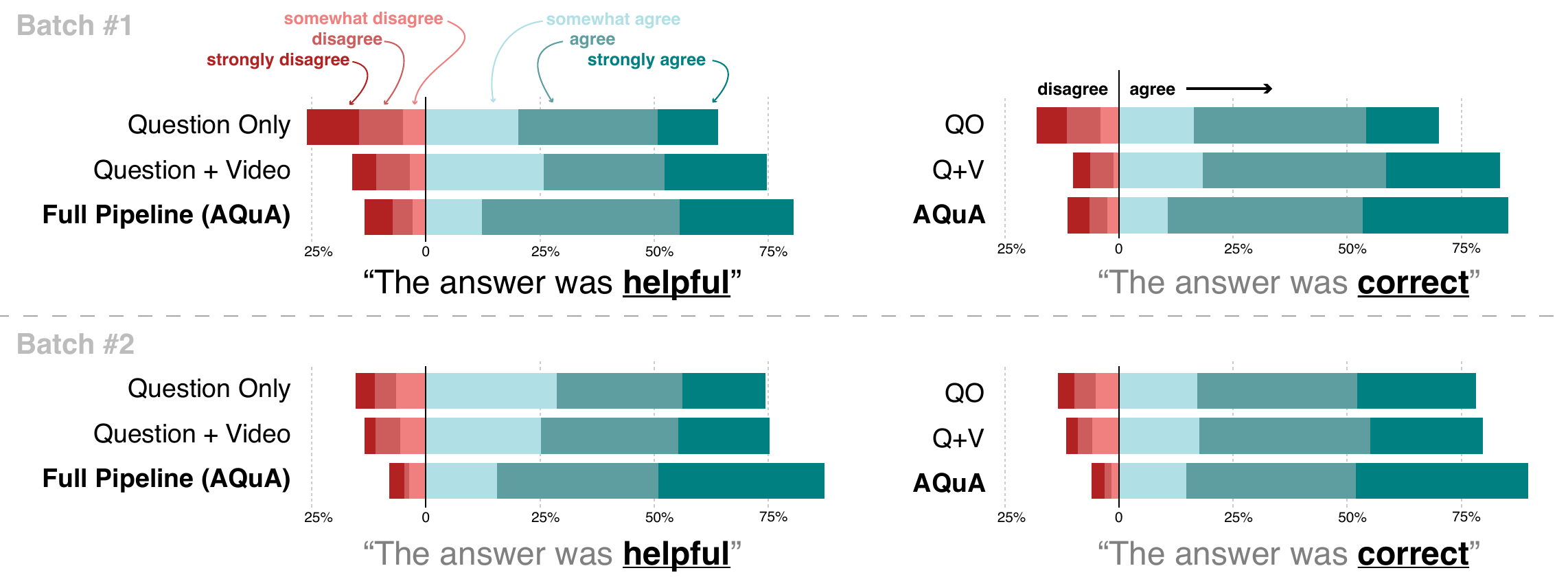}
    \caption{Distribution of Likert scale responses on Correctness and Helpfulness. \texttt{Full Pipeline} shows the highest correctness and helpfulness scores in both batches. Responses of "neither agree nor disagree" are omitted from the chart for clarity and readability.}
  \label{fig:ratings}
  \Description{Bar graphs that show the result of Likert scale responses on Correctness and Helpfulness for each batch.}
\end{figure*}

\begin{figure}[t]
    \centering
    \includegraphics[width=\linewidth]{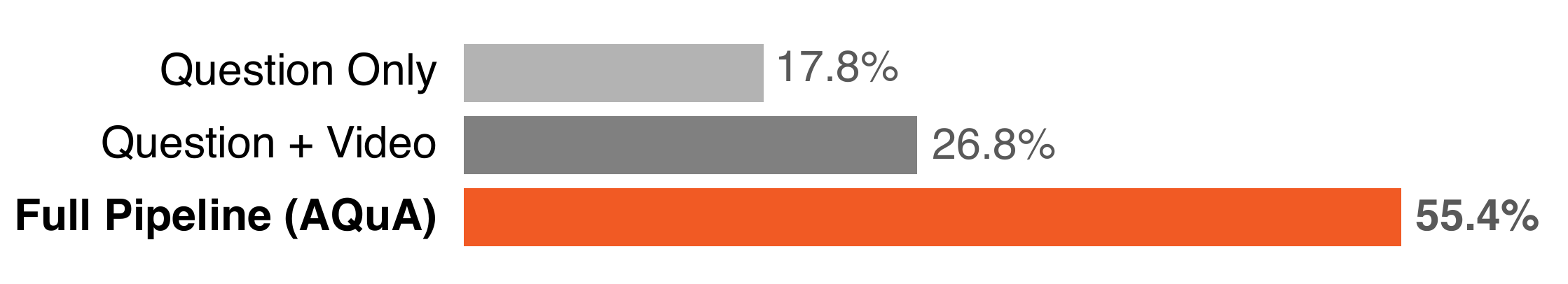}
    \caption{Results of the favorite answer selection. Answers generated from the \texttt{Full Pipeline} were selected as the favorite most often.}
    \label{fig:favorite}
    \Description{Bar graph that shows the result of favorite answer selection. Question Only: 17.8\%, Question + Video: 26.8\%, Full Pipeline: 55.4\%.}
% \vspace*{6pt}
\end{figure}

\subsubsection{Full Pipeline Favored Over Baseline Answer Generation Methods}
In selecting a preferred answer, the answer generated by \texttt{Full Pipeline} was favored most frequently (55.4\%), in comparison to the \texttt{Question-only} (17.8\%) and \texttt{Question + Video} condition (26.8\%), as shown in Figure~\ref{fig:favorite}. The trend remained consistent for both Batch 1 and Batch 2 -- 15.8\%, 29.4\%, and 54.8\% for Batch 1 and 19.6\%, 24.3\%, and 56.1\% for Batch 2, corresponding to \texttt{Question-only}, \texttt{Question + Video}, and \texttt{Full Pipeline}, respectively. In the open-ended feedback in which participants could provide feedback on why they selected \texttt{Full Pipeline} answers as their favorite, they noted that these answers were both accurate and specific (P4, P12, P14, P15, P16), provided more detail than others (P13, P14, P16), and followed the right sequence of addressing a question (P8, P16). For instance, in the top example from Table~\ref{table:answer_examples}, \texttt{Question-only} failed to comprehend the specific menu referred to in the question and \texttt{Question + Video} incorrectly inferred that it related to another tool based on the transcript instructions. However, \texttt{Full Pipeline} accurately identified the menu as the Fusion~360 \emph{Marking Menu} and provided precise information about it. Additional examples of answers can be found in Table~\ref{table:answer_examples}.

We also conducted a Friedman test to examine whether differences existed between the answer conditions. Since Batch 1 and Batch 2 were evaluated by separate sets of users, we performed the analysis independently for each batch. We observed statistically significant differences between answer conditions in both batches (Batch 1: ${\chi}^2$ = 48.5, \textit{p} < .001 and ${\chi}^2$=43.4, \textit{p} < .001, Batch 2: ${\chi}^2$=32.5, \textit{p} < .001 and ${\chi}^2$=62.8, \textit{p} < .001 for helpfulness and correctness, respectively). 
Subsequently, we conducted post-hoc analysis with a Nemenyi test to identify which groups accounted for these differences. For both batches, the answers generated by \texttt{Full Pipeline} were significantly more correct and helpful compared to the two other methods (p = .009 for \texttt{Question + Video} vs. \texttt{Full Pipeline} in Batch 1, p = .001 for the rest).  
Additionally, the correctness of \texttt{Question + Video} surpassed that of \texttt{Question-only} in Batch 1 (\textit{p} = .013). Figure~\ref{fig:ratings} displays the distribution of Likert scale responses for each condition across both batches.

\subsubsection{When Did the Full Pipeline Fail?}
There were instances where \texttt{Full Pipeline} fell short in generating useful answers compared to the other conditions. It sometimes assembled unrelated information, making the answers unnecessarily complicated. For instance, when asked about the \texttt{nesting sketches} feature in Fusion 360, P8 noted that the \texttt{Full Pipeline}'s answer mixed the concepts of nesting sketches with nesting for manufacturing. Similarly, P13 pointed out that it was overly verbose and used complex language, making it difficult to understand. These observations suggest that retrieving the right amount of information is crucial for providing useful answers. Additionally, there were a few instances in which it generated incorrect information by not recognizing UI elements in visual anchors that were not present in the UI database. In that case, the pipeline had to rely on other available information, such as transcripts, which led to incorrect answers.

\subsubsection{General Feedback on AI-generated Questions}
Overall, participants felt that the answers offered fairly accurate information (P1, P2, P7, P14, P15). P2 remarked that these AI-generated answers have the potential to provide quick and clear responses, thereby eliminating the need to sift through lengthy videos, forum threads, or posts. Despite these strengths, areas for improvement were identified. P16 suggested enhancing the terminology and style used in the answers, following the guidelines used in the software. P15 recommended that answers be more direct and to the point. For instance, mentioning ``Fusion 360'' in an answer is unnecessary as it is already evident. We discuss design guidelines for question-answering systems in Section~\ref{sec:guidelines}.

\section{Discussion and Future Work}
In this section, we discuss design considerations for ques\-tion-an\-swer\-ing systems, possible question-answer pipeline improvements, potential video question-answering interface designs built on top of our pipeline, and generalizability to other \nr{software and} domains.

\subsection{Design Considerations for Question-Answering Systems}\label{sec:guidelines}
\pointsix{Providing answers to users' questions as they learn and follow software tutorial videos is crucial. For instance, it can offer personalized explanations that go beyond one-size-fits-all tutorials, by addressing points of confusion unique to different learners. Additionally, it can guide users on resolving current challenges they encounter while following tutorials, a common issue faced by learners~\cite{yang2020hard}.}
We identified a number of factors that are important to consider in the design of automated question-answering systems from our evaluation study, which we discuss below.

\subsubsection{The Right Tone}
Maintaining an appropriate tone in answers is key. For instance, there was a case where a user noted a discrepancy between the provided instruction and the outcome after following it. They questioned what might have gone wrong, and P16 felt that the responses generated subtly laid blame on the user (e.g., ``It's possible that a step was missed or misunderstood''). However, P16 appreciated one answer that acknowledged the complexity of Fusion 360, implicitly allowing for the possibility that the instructions might not have been easy to follow. This led P16 to select that response as the preferred one, underscoring the importance of tone in responses. As a system assisting the user, it's essential to avoid attributing blame and to provide help in a constructive manner.

\subsubsection{Answer Length}
LLMs like ChatGPT often generate lengthy answers, which might be informative but not always ideal for users seeking brief, straightforward information~\cite{guo-etal-2023-hc3}. In our first formative study, we observed that human-generated answers typically consist of 1 to 2 sentences. Accordingly, we instructed GPT-4 to limit responses to 50 words or fewer in all three conditions. However, our evaluation study still revealed a divide within the word limit: some participants preferred detailed answers for practical instruction, while others sought concise responses, believing that excess details could obscure the main point. Thus, an ideal approach would need to consider the user's preferences and prior knowledge when formulating responses. A strategy that might work for a diverse audience is offering expandable answers: beginning with a concise response and allowing users the option to see more details, or alternatively, asking the system to elaborate on its answer (see also Section~\ref{sec:discussion-conversational}).

\subsubsection{Transparency about Uncertainty}
LLMs are susceptible to generating false or misleading information, known as hallucinations~\cite{zhang2023hallucination}. In our evaluation study, a number of participants pointed out that an answer mentioned unrelated tools (P14) or unavailable operations (P2). While experienced users might identify such inaccuracies, new users could easily be misled. Therefore, it is crucial for LLMs to be transparent about their limitations and uncertainties. Recent work on explainable LLMs, such as the ability to cite specific evidence for claims~\cite{menick2022teaching}, could be beneficial in this context. Additionally, the system could be designed to be more interactive. For example, if the system fails to comprehend the question or identify a visual anchor referenced in it, it should prompt the user for additional details, enabling more context-rich and accurate answers.

\begin{figure*}[h]
  \includegraphics[width=\linewidth]{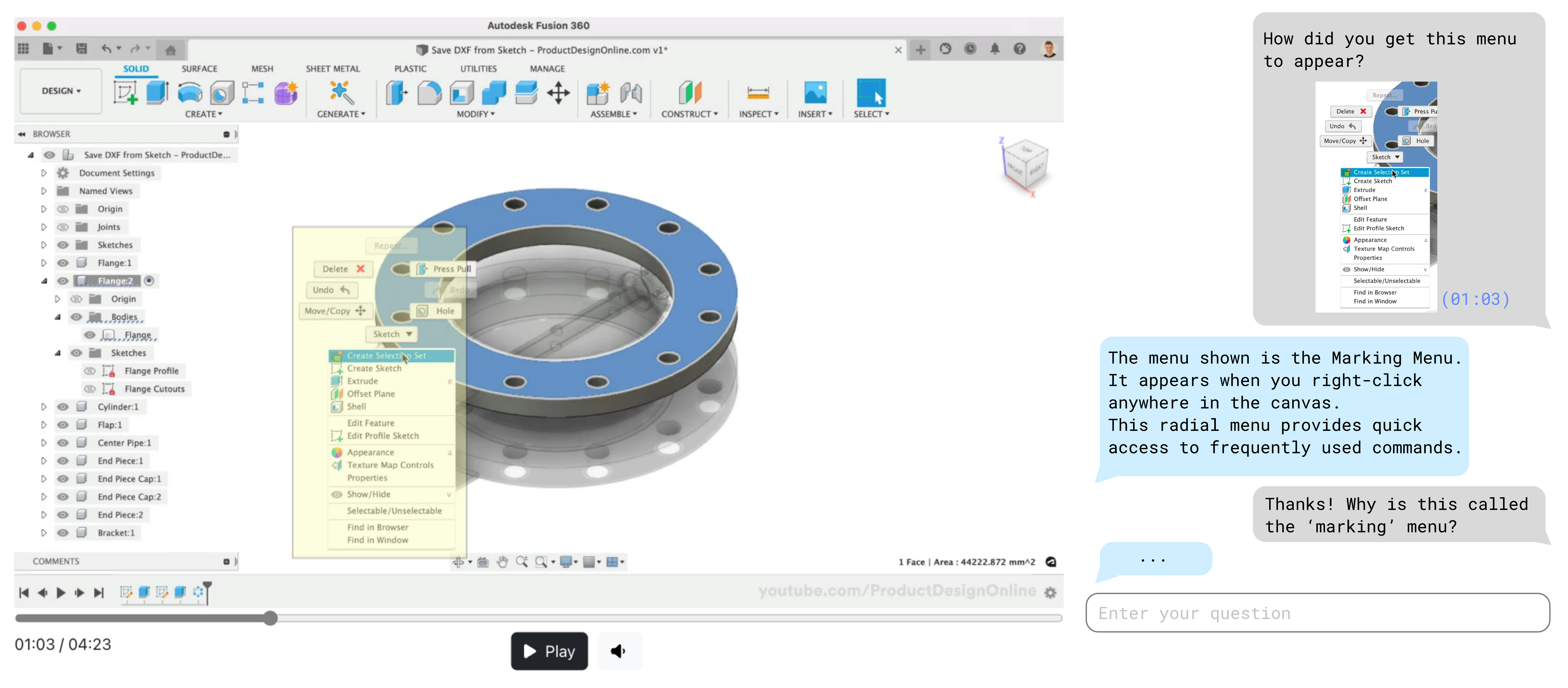}
  \caption{We envision that our pipeline could be leveraged in the future to develop a tutorial video system that supports conversational, chat-like question and answering. Learners could ask questions by referring to specific parts of the video. The system would then generate responses to these questions, while also allowing users to easily ask follow-up questions.}
  \Description{A system that shows a tutorial video on the left. The user has drawn a visual anchor on top of the video screen. On the right, the user is having a text chat conversation with the system by asking questions and getting answers.}
  \label{fig:envision}
\end{figure*}

\subsection{Improvements to the Question-Answer Pipeline}
% \rr{\emph{AQuA} is designed to handle questions related to tutorial (`Content' and `User'). In our evaluation, 13\% of the questions were `User' questions }
\pointfive{\emph{AQuA} recognizes visual anchors, retrieves relevant software-specific materials, and includes tutorial video context to generate answers to a given question. 
This multimodal approach allows for a more comprehensive understanding of the user's query and thereby generates accurate and helpful answers (Section~\ref{sec:evaluation}).
However, achieving this involves several components working together in the pipeline. Since these modules run sequentially, there is a potential for latency issues. In our case, answer generation takes around 30 seconds once the offline models, such as BLIP-2 and GPT-4, are loaded. 
We believe future work could explore ways to further reduce latency, enabling nearly real-time question-answering.}

% While we found that our pipeline generates accurate and helpful answers overall (Section~\ref{sec:evaluation}), there are a number of areas in which it could be further enhanced.
\rr{Apart from the latency, there are a number of areas in which our pipeline could be further enhanced.}
First, the pipeline could be extended to identify the precise point of interest within a given visual anchor. Participants sometimes captured a larger area in their visual anchors than the specific subject of their question, and sometimes, the precise visual anchor could not be derived from the question text. Although visual anchors already provide a narrower scope compared to a full video frame, a further scope reduction could yield more accurate results. This specific point of interest could be inferred from the mouse pointer's location while creating the visual anchor. Alternatively, the Visual Recognition Module could be integrated into an intuitive interface to allow users to select a specific visual anchor from a set of automatically recognized elements of interest in the video frame.
Second, the pipeline's robustness could be increased by incorporating additional resources. Relevant content from other online tutorials, Q\&A forums, and previous user comments could be integrated to provide a more comprehensive answer.
Third, to offer richer context for the tutorial video in question, the system could go beyond using the title and transcript for context. By employing a video captioning model specifically trained on screencast videos, such as the one by Li et al.~\cite{li2020pstuts}, we could obtain a more nuanced understanding of the video's content at the time the question was posed.
\rr{Lastly, if we take into account learners' progress on their software as in SoftVideo~\cite{yang2022softvideo}, we can offer more specific and detailed answers tailored to users' individual levels of knowledge or current progress. Understanding users' preferences and proficiency levels with the software can be especially useful when answering questions in the `User' category.
}

% Potential Video Question-Answering Interface Designs
\subsection{\sr{Potential Interface Designs for Tutorial Video Systems}}

Our work opens up exciting opportunities for integrating the ques\-tion-an\-swer\-ing pipeline into a tutorial video system. Here, we discuss a number of interface ideas that we find particularly promising.

\subsubsection{Conversational Question-Answering}\label{sec:discussion-conversational}
Beyond single-turn ques\-tion-an\-swer\-ing, multi-turn conversations can facilitate deeper understanding and assistance. 
We envision that a promising interface design for our \emph{AQuA} pipeline is a conversational, chat-based question-answering system similar to ChatGPT~\cite{chatgpt}, as illustrated in Figure~\ref{fig:envision}. This would allow users to ask follow-up questions or provide feedback on the answers they receive, which the system can incorporate into future answers.
Moreover, the system could even proactively initiate conversations to monitor users' progress and assess their comprehension\ar{, as in Shin et al.~\cite{shin2018inprompt},} thus delivering more personalized responses. The system could also include motivational phrases such as, ``You're asking great questions!'' or ``That's correct, you're a fast learner!'' to further inspire and engage learners~\cite{alcott_2017}.
% Encouragement also has a meaningful impact on learners' progress.
We believe these enhancements offer valuable opportunities for effective learning experiences with tutorial videos.

\subsubsection{Support for Transcript Anchors}
Our pipeline provides useful answers to queries that include visual anchors. An interesting interface extension could be to allow users to not only reference visual elements of interest, but also refer to parts of the audio transcript, which is another common type of reference in video~\cite{yarmand2019reference}. Users could select or drag over parts of the transcript and ask questions about it (an example for Fusion~360 could be: \textit{``What do you mean by `reference a construction plane', and how is that done?''}). Allowing users to refer to both elements in the video and in the transcript can help them better articulate their questions. 

\subsubsection{Making Video Comments More Useful}
Our approach to supporting questions and comments with visual anchors also opens up potential improvements to the interface design of (tutorial) video interfaces. 
%First, it allows users to easily access previous learners' questions (as well as answers to these questions) in the context of following a tutorial video. These questions and answers could be surfaced progressively as the user advances through the video. 
%Questions with visual anchors could be used to organize user comments in a more structured way. While user comments can provide a wealth of insights relevant to the video~\cite{Madden2013ACS}, 
Traditional video interfaces often separate video content from user comments, making it challenging to locate relevant discussions. 
With our approach, comments and automated answers can be organized based on the visual objects or components appearing in the video. For example, a user could visually select a tool of interest that is featured in the video to see related questions and comments about that tool. 
\pointsix{This could also be advantageous for tutorial authors by offering a quick overview of areas that generate the most questions, as demonstrated in Mudslide~\cite{mudslide}. By efficiently reviewing questions from learners, authors can identify areas of confusion or topics that require more elaboration. These insights can serve as valuable feedback for authors when creating the next tutorial video. Furthermore, an interesting direction could be to simulate learners' behavior, as explored in Generative Agents~\cite{park23agents}, and generate simulated questions even before publishing the video. This would enable authors to enhance their tutorial content by addressing potential points of clarification in advance.}

% Lastly, it can also introduce new means of video navigation by allowing users to navigate the video based on visual objects~\cite{direct-object}. Users can indicate an object of interest, similar to creating visual anchors. The system can then locate segments where these objects appear throughout the video,  making it easier for users to find relevant content.

\subsection{\pointsixcolor{Generalizability to Other Software}}
As discussed in Section~\ref{sec:pipeline}, we believe our question-answer pipeline \emph{AQuA} can easily generalize to other feature-rich software applications, such as Photoshop or AutoCAD. 
\pointsix{
To adapt our approach to different software, only two components require replacement: (1) the \emph{UI database}, encompassing software icons and names, and (2) \emph{software articles}, such as documentation or tutorials. These resources are designed to recognize software UI elements in the visual anchor and provide software-specific information. In our demonstration with Fusion 360, we constructed these databases by crawling publicly available sources (details in Section~\ref{sec:pipeline}). This implies the possibility of constructing similar databases for other software applications using their official documentation and publicly available tutorial resources. The remaining components would work the same, and by leveraging an off-the-shelf pre-trained LLM, we minimize the need for additional computational resources when adapting to other software applications. This approach makes our pipeline extensible and facilitates adapting the pipeline to various other software applications in the future.
}

\subsection{Generalizability to Other Domains}
It would be interesting to explore expanding the scope of our approach to other learning domains, such as instructional videos that teach physical skills\pointsix{, programming tutorials, or lecture videos}. These videos, much like software tutorials, often \mr{convey} information through both visual and verbal channels~\cite{democut, stargazer,shin15visual}, which suggests the potential for questions with visual anchors. 
Given that LLMs and image captioning models are well-equipped with knowledge associated with everyday tasks and objects such as cooking and assembling furniture, it is conceivable that \emph{AQuA}'s capabilities could also extend to these domains, by using different resources (e.g., recipes and cookbooks instead of software documentation). For instance, users could anchor a question to a specific ingredient in a cooking video to inquire about its function and possible substitutes. \pointsix{The image captioning models are able to recognize ingredients, and by leveraging the rich knowledge encompassed by LLMs and enhanced by recipe-specific resources, the pipeline would likely be able to provide a comprehensive answer. On the other hand, for programming or lecture videos, we could rely more on OCR results as these videos often contain text-heavy content. Together with knowledge already embedded in LLMs and leveraging more specific materials such as textbooks, our pipeline could likely offer accurate answers.}
We believe that \mr{with some adjustments}, our question-answering system with support for visual anchors could enable more contextual and comprehensive help systems across \mr{various} domains.

% \emph{AQuA} will work for any software application for which tool/command names accompanied by icons or screenshots of corresponding UI elements, software documentation with descriptions of tools and their purpose, and a sufficiently large data set of existing tutorial videos are available. This is, for example, be the case for other feature-rich software like Adobe Photoshop.

\section{Conclusion}
We introduced an automated approach for answering questions in software tutorial videos. To achieve this, we conducted two formative studies to understand users' question-asking behavior. \mr{Focusing on questions related to the tutorial content,} we discovered that users frequently refer to visual elements of the video, particularly focusing on UI components and the application workspace. Based on these insights, we developed \emph{AQuA}, \mr{an LLM-based multimodal} pipeline that generates useful answers to questions that include visual anchors, which are specific visual elements of interest in the tutorial video. Using software-specific resources such as software documentation and icons of tools, our pipeline identifies these visual anchors and generates answers tailored to the particular software. Our evaluation demonstrated that our approach yields more accurate and more helpful responses compared to baseline methods. Lastly, we discuss design considerations for question-answering systems and promising directions for future work, offering insights into the future of interactive and responsive learning experiences.
\begin{acks}
We thank Amir H. Khasahmadi, Michael Chen, and Fraser Anderson from Autodesk Research for their help and feedback.
% Should thank Amir (screencast data), Michael (Beware), and Fraser (stats) for help. We don't need to go into detail about what they did, but can just list their names for help/feedback.
\end{acks}

\bibliographystyle{ACM-Reference-Format}
\bibliography{references}

\appendix
\section{Prompts Used in Question-Answer Pipeline}\label{sec:prompts}
\subsection{(1) Question-Only}\label{sec:prompts_no}
% \begin{lstlisting}[breaklines=true, breakatwhitespace=true]
\begin{lstlisting}
You need to answer questions about Autodesk Fusion 360 that people asked while watching a tutorial video. Please answer in 50 words or less. 

Question: {question_text}

\end{lstlisting}

\subsection{(2) Question and Video Context}\label{sec:prompts_baseline}
% \begin{lstlisting}[breaklines=true, breakatwhitespace=true]
\begin{lstlisting}
You need to answer questions about Autodesk Fusion 360 that people asked while watching a tutorial video. Please answer in 50 words or less. 

Tutorial: Title: {title}. Instructions: {transcript}
Question: {question_text}

\end{lstlisting}

\subsection{(3) Our Full Pipeline}\label{sec:prompts_full}
% \begin{lstlisting}[breaklines=true, breakatwhitespace=true]
\begin{lstlisting}
You need to answer questions about Autodesk Fusion 360 that people asked while watching a tutorial video. Please answer in 50 words or less. Each question is accompanied by relevant visual anchors, which are specific visual elements of interest in the video.


Use the below articles on the Fusion 360 software to answer the subsequent question. If the answer cannot be found in the articles, write ``I could not find an answer.''
Fusion 360 article section: {section 1}
Fusion 360 article section: {section 2}
...

Tutorial: Title: {title}. Instructions: {transcript}
Question: {question_text}
Visual Anchor: 
{Anchor_label_1}: {blip}. It includes the Fusion 360 tools: {tool} and text: {ocr}.
{Anchor_label_2}: {blip}. It includes the Fusion 360 tools: {tool} and text: {ocr}.
...

\end{lstlisting}

\end{document}